\documentstyle[epsfig,preprint,pra,aps]{revtex}

\begin{document}
\title{\bf $ $ \\ Andreev Bound States and Self-Consistent
Gap Functions for SNS and SNSNS Systems }

\author{R. E. S. Otadoy\footnote{On leave of absence
      from the University of San Carlos, Nasipit, Talamban, Cebu City,
      Philippines}  and A. Lodder\footnote
        {Corresponding author, e-mail address: alod@nat.vu.nl}\\
Faculteit Natuurkunde en Sterrenkunde, Vrije Universiteit De
Boelelaan 1081, \\ 1081 HV Amsterdam, The Netherlands}

\date{\today}
\maketitle

\begin{abstract}
\normalsize{
Andreev bound states in clean, ballistic SNS and
SNSNS junctions are calculated exactly and by using the Andreev
approximation (AA). The AA appears to break down for junctions
with transverse dimensions chosen such that the motion in the
longitudinal direction is very slow. The doubly degenerate states
typical for the traveling waves found in the AA are replaced by
two standing waves in the exact treatment and the degeneracy is
lifted.

A multiple-scattering Green's function formalism is used, from
which the states are found through the local density of states.
The scattering by the interfaces in any layered system of
ballistic normal metals and clean superconducting materials is
taken into account exactly. The formalism allows, in addition, for
a self-consistent determination of the gap function. In the
numerical calculations the pairing coupling constant for aluminum
is used. Various features of the proximity effect are shown.
}
\end{abstract}

\section{Introduction}

In 1991, Tanaka and Tsukada \cite{Tanaka} investigated the energy
spectrum of the quasiparticle states in a superconducting
superlattice based on a Kronig-Penney model potential. The
calculations were done using the Andreev approximation
\cite{Andreev} and the system considered was infinite in both the
transverse and the longitudinal dimensions. In this paper we
concentrate on three additional aspects: \textit{i}) we do the
exact calculation and investigate the reliability of the Andreev
approximation, \textit{ii}) we do the calculation for a limited
number of layers, which relates better to a possible experimental
situation, and \textit{iii}) we investigate the transverse size
dependence of the relevant properties such as the local density
of states. The systems we consider cover the entire range from
very narrow transverse dimensions to wider ones, thereby
simulating 1, 2 and 3-dimensional systems.

The authors are aware of two studies focusing on the transverse
size dependence of Andreev bound states and the quasiparticle
local density of states. \v{S}ipr and Gy\"{o}rffy \cite{Gyorffy}
show the splitting of the degenerate bound states found in the
Andreev approximation by studying the exact solution of the
Bogoliubov-de Gennes equations. Their study is limited to a
superconductor-normal metal-superconductor (SNS) junction and no
self-consistency of the gap is considered. A second study, by
Blaauboer et al. \cite{Miriam}, is a rather global one, but in
that study the multiple-scattering Green's function formalism
developed by Koperdraad \cite{Koperdraad} was applied for the
first time. The local density of states of normal
metal-superconductor (NS) and SNS junctions were calculated, but
the breakdown of the Andreev approximation was not noticed.

We want to extend these studies by considering self-consistently
obtained gap functions and by studying a system with more than
two interfaces, looking for the influence of an additional
superconducting layer to the Andreev bound state spectrum. The
multiple-scattering Green's function formalism to be used
\cite{Koperdraad} is an extension of the formalism introduced by
Tanaka and Tsukada \cite{Tanaka} in that no Andreev approximation
is made. In addition it is set up in a quite  different way, such
that the multiple scattering of the (quasi)particles by the
different interfaces is exhibited explicitly.

The paper is organized as follows. First we give a concise
account of the main features of the theory. In Sec.~\ref{LDOS} the
local density of states for the two systems considered is
studied. In Sec.~\ref{gap} the gap function is calculated
self-consistently for a bar-shaped superconductor and for the NS,
SNS and SNSNS junctions. Concluding remarks are given in
Sec.~\ref{Conclusion}.

Throughout the paper, Rydberg atomic units are used, in which the
energy is in Rydberg, the distance is in Bohr ($1$ Bohr $\approx$
0.5 \AA), $\hbar=1$, and the electronic mass is $\frac{1}{2}$.

\section{Theory}
\label{theory}

Although in this paper a Green's function formalism is used, we
first give the Bogoliubov-de Gennes equations
\begin{eqnarray}
\label{Bogol1} \left[ \begin{array}{cc}
  - {\nabla}^2 - \mu &   \Delta({\bf r}) \\
 \Delta^{\ast}({\bf r}) &
 {\nabla}^2 + \mu
\end{array}\right]
\Psi({\bf r}) = E \Psi({\bf r})\equiv E
\left(\begin{array}{c}u({\bf r}) \\
 v({\bf r}) \end{array}  \right)
\end{eqnarray}
used by \v{S}ipr and Gy\"{o}rffy \cite{Gyorffy} to investigate the
bound states of a SNS junction shown in the upper panels of
FIG.~\ref{SNSNSfig}. The main reason is that our formalism will
be expressed in terms of the solutions of these equations. The
inhomogeneity of the system is expressed by the space dependence
of the gap $\Delta({\bf r})$. In the superconducting parts the
gap is a complex constant whereas in the normal metal it is zero.
The spinor wavefunction describes quasiparticle excitations and
the energy $E$ is measured with respect to the Fermi energy $\mu$.

Application of the Bogoliubov-de Gennes equations to the
bar-shaped superconductor  shown in FIG.~\ref{bar} yields
\begin{equation}
\label{solution1}
 \Psi({\bf r})=
 \left( \begin{array}{c}
 u_S^{\sigma} e^{i\phi/2} \\ u_S^{-\sigma} e^{-i\phi/2}
 \end{array} \right)
 e^{i\sigma\nu k_S^{\sigma} x}
 \sin \left( \frac{n_y\pi}{L_y} y \right)
 \sin \left( \frac{n_z\pi}{L_z} z \right),
\end{equation}
where $\phi$ is the phase of the complex constant $\Delta$, and
\begin{eqnarray}
\label{transwavevector} k_y&=&\frac{n_y\pi}{L_y}\;;\;
k_z=\frac{n_z\pi}{L_z} \hspace{5mm}{\rm with}
\hspace{3mm}n_y , n_z =1,2,3,\cdots\\
\label{usigma}
u_S^{\sigma}&=&\sqrt{E + \sigma \sqrt{E^2 - |\Delta|^2}}\\
\label{ksigma}
k_S^{\sigma}&=&\sqrt{k_{F_x}^2 + \sigma\sqrt{ E^2 - |\Delta|^2}}\\
\label{kFx}
k_{F_x}^2&=&\mu-k_y^2-k_z^2
\end{eqnarray}
The discretized nature of the transverse wavevectors is due to
the vanishing of the wavefunctions on the transverse boundaries.
The four standard solutions are labeled with the sign indices
$\sigma$ and $\nu$ that can both be equal to $\pm1$. By this
convention, the index $\sigma$ refers to the type of particle
(electronlike for $\sigma=+1$ and holelike for $\sigma= -1$) and
the index $\nu$ indicates the direction of propagation ($\nu=+1$
to the right and $\nu=-1$ to the left). The complete solution of
Eq. (\ref{Bogol1}) is a linear superposition of Eq.
(\ref{solution1}) for different possible combinations of $(\sigma
, \nu)$. The above equations also hold for a normal metal by
letting $\Delta=0$, in which case the subscript $S$ is to be
replaced by $N$.

The Green's function formalism is outlined extensively by
Koperdraad et al. \cite{Koperdraad}. It is an extension of the
microscopic theory used by Tanaka and Tsukada, in that the
electron-hole scattering properties are treated exactly, and it
traces back to the microscopic description of superconductivity
by Gor'kov \cite{Gorkov} and Ishii \cite{Ishii}. For the sake of
completeness and clarity we summarize its main features and add
relevant new elaborations. The matrix Green's function satisfies
the equation
\begin{equation}
\label{Green1}
 \left[\begin{array}{cc}
  i\omega_n+{\nabla}^2+\mu & -\Delta({\bf r})\\
  -\Delta^{\ast}({\bf r}) & i\omega_n-{\nabla}^2-\mu
 \end{array}\right]G({\bf r},{\bf r'},i\omega_n)
 =\delta({\bf r}-{\bf r'})\mathbf{1}
\end{equation}
in which the differential operator is closely related to the
operator in the Bogoliubov-de Gennes equations (\ref{Bogol1})
apart from the replacement of $E$ by $i\omega_n$ where
\begin{equation}
\label{matsubara} \omega_n=\pi nk_B T ~{\rm with}~n~{\rm a~\pm
~odd~integer}.
\end{equation}
The quantity $\omega_n$ is called the Matsubara frequency.
Possible inhomogeneities of the system are fully represented by
the $\mathbf{r}$ dependence of the gap function. As far as the
transverse directions are concerned the general solution of Eq.
(\ref{Green1}) can be expressed as a linear combination of
solutions (\ref{solution1}) over all allowed values of
$k_y,k'_y,k_z,$ and $k'_z$. Thus
\begin{equation}
\label{Fourier} G({\bf r,r'},i\omega_n)=\frac{4}{L_y L_z}
\sum\limits_{k_y,k'_y,k_z,k'_z}G(x,x',k_y,k'_y,k_z,k'_z,i\omega_n)
\sin k_y y\sin k'_y y'\sin k_z z\sin k'_z z'.
\end{equation}
The transverse standing waves are normalized to unity, which
implies the orthogonality condition
\begin{equation}
\label{orthogonality} \int_0^{L_y}\sin k_y y\sin \bar{k}_y y dy
=\frac{L_y}{2} \delta_{k_y,\bar{k}_y},
\end{equation}
for the y coordinate and a similar condition for the z coordinate.
The Fourier coefficient $G(x,x',k_y,k'_y,k_z,k'_z,i\omega_n)$ can
be obtained from Eq. (\ref{Fourier}) using Eq.
(\ref{orthogonality}). It takes the form,
\begin{eqnarray}
G(x,x',k_y,k'_y,k_z,k'_z,i\omega_n)&=&\frac{4}{L_y
L_z}\int_0^{L_y}\sin k_y y \int_0^{L_y}\sin k'_y
y'\int_0^{L_z}\sin k_z z \int_0^{L_z}\sin k'_z z'\nonumber\\
\label{Fouriercoef}
& &\times G({\bf r,r'},i\omega_n)dz'dzdy'dy.
\end{eqnarray}
Using Eq. (\ref{Fouriercoef}), Eq. (\ref{Green1}) can be written
as
\begin{equation}
\label{Green2} \left[\begin{array}{cc}
i\omega_n+\frac{d^2}{dx^2}+k_{F_x}^2 & -\Delta \\
-\Delta^{\ast} &
i\omega_n-\frac{d^2}{dx^2}-k_{F_x}^2\end{array}\right]
G(x,x',k_y,k'_y,k_z,k'_z,i\omega_n)
=\delta(x-x')\delta_{k_y,k'_y}\delta_{k_z,k'_z}{\bf 1}
\end{equation}
Before we proceed we want to make the terminology clear. $G({\bf
r,r'},i\omega_n)$ is the actual Green's function. The Fourier
coefficient $G(x,x',k_y,k'_y,k_z,k'_z,i\omega_n)$ is a Green's
function in the sense that it is the solution of Eq.
(\ref{Green2}), but this equation demonstrates that it is diagonal
in $k_y$ and $k_z$. That facilitates both the notation, because
the variables $k'_y$ and $k'_z$ can be omitted, and the
calculation of $G({\bf r,r'},i\omega_n)$ according to Eq.
(\ref{Fourier}). In calculating the local density of states and
the self-consistent gap function, we will need the Green's
function for diagonal spatial coordinates. As long as we keep
$x\neq x'$, we can already take $y=y'$ and $z=z'$ in Eq.
(\ref{Fourier}). Since we are only interested in the variations
over the longitudinal direction $x$, we can average over the
transverse dimensions. By that Eq. (\ref{Fourier}) simplifies to
the series
\begin{equation}
G(x,x',i\omega_n)=\frac{1}{L_y
L_z}\sum\limits_{k_y,k_z}G(x,x',k_y,k_z,i\omega_n).
\end{equation}

First we give the Green's function of a homogeneous bar-shaped
superconductor
\begin{equation}
\label{Green3}
 G_S^0(x,x',k_y,k_z,i\omega_n)=\sum_{\sigma}d_S^{\sigma}\psi_S^{\sigma-}(x_<)
 \tilde{\psi}_S^{\sigma +}(x_>)=\sum_{\sigma}d_S^{\sigma}
\psi_S^{\sigma +}(x_>)\tilde{\psi}_S^{\sigma -}(x_<), ~{\rm with}
~d_S^\sigma=-\frac{1}{4\Omega_S k_S^\sigma},
\end{equation}
in which $i\Omega_S=\sqrt{(i\omega_n)^2 - |\Delta|^2}$. The
wavefunction $\psi_S^{\sigma\nu}\left(x\right)$ is given by Eq.
(\ref{solution1}) after omitting the transverse solutions, and
$u_S^{\sigma}$ and $k_S^{\sigma}$ are given by Eqs.
(\ref{usigma}) and (\ref{ksigma}) with the replacement $E\to
i\omega_n$. In addition the conjugate wavefunction is defined by
\begin{equation}
\label{psiconjug}
 \tilde{\psi}_S^{\sigma\nu}(x)=
 \left(
   \begin{array}{cc}
    u_S^{\sigma}e^{-i\phi/2} & u_S^{-\sigma}e^{i\phi /2}
   \end{array}
 \right)
 e^{i \sigma \nu{k_S^{\sigma} x}}
\end{equation}
The Green's function (\ref{Green3}) describes the propagation of
excitations (holelike or electronlike) from the starting point
$x'$ to the final point $x$ with the weighting factor
$d_S^{\sigma}$. In this case of a superconducting bar, there is no
scattering.

Now we give the Green's function for a system with one interface.
To account for the scattering at the interface, its appropriate
form appears to be
\begin{equation}
\label{GreenNS} G_{\nu j\nu 'j}(x,x',k_y,k_z,i\omega_n)= G_{\nu
j}^0 (x,x',k_y,k_z,i\omega_n)\delta_{\nu \nu '}
  +\sum_{\sigma\sigma '}d_{\nu j}^{\sigma}d_{\nu 'j}^{\sigma '}
  \psi_{\nu j}^{\sigma\nu}(x)t_{\nu j\nu 'j}^{\sigma \sigma ' \nu \nu '}
  \tilde{\psi}_{\nu 'j}^{\sigma ' \nu '}(x')
\end{equation}
where
\begin{equation}
\label{auxGreenNS}
  G_{\nu j}^0 (x,x',k_y,k_z,i\omega_n)=\sum_{\sigma}d_{\nu j}^{\sigma}
 \psi_{\nu j}^{\sigma \mu}(x)\tilde{\psi}_{\nu 'j}^{\sigma ' ,-\mu}(x')
 ~{\rm with}~\mu=\mathrm{sgn}(x-x').
\end{equation}
The label $S$ in Eq. (\ref{Green3}) has been replaced with the
more flexible label $\nu j$ indicating the position $x_j$ of the
interface. The subscript $+j$ means $x$ is in the right-hand side
of the interface $x_j$ and $-j$ means that it is in the left-hand
side. The first term is nonzero only if the starting and final
points, $x'$ and $x$ respectively, are at the same side of the
interface. The second term takes into account the scattering of
the particle at the interface. This scattering is aptly described
by the parameter we call scattering $t$-matrix, $t_{\nu j\nu
'j}^{\sigma\sigma '\nu\nu '}$. For a superconducting bar the
$t$-matrix is zero, because there is pure propagation and no
scattering. The scattering $t$-matrix can be obtained by imposing
the continuity of the Green's function and its derivative at the
interface. We then obtain
\begin{equation}
\label{tmatrix}
 \sum_{\sigma\nu}{\nu}d_{\nu j}^{\sigma}\psi_{\nu j}^{\sigma \nu}
 (x_j)t_{\nu j\nu 'j}^{\sigma\sigma '\nu\nu '}=-\nu '
 \psi_{\nu 'j}^{\sigma '\,-\nu '}(x_j).
\end{equation}
In applying the above interface matching condition, it has come
out to be convenient to use an extended definition
\begin{equation}
\label{solution6}
 \psi_S^{\sigma\nu}(x)=\left(\begin{array}{c}
 u_S^{\sigma}e^{i\phi/2}\\u_S^{-\sigma}e^{-i\phi/2}
 \\i{\sigma\nu}k_S^{\sigma}u_S^{\sigma}e^{i\phi/2}
 \\i{\sigma\nu}k_S^{\sigma}u_S^{-\sigma}e^{-i\phi/2}
\end{array}\right)e^{i{\sigma\nu}k_S^{\sigma}x}
\end{equation}
of the wavefunction $\psi_S^{\sigma\nu}\left(x\right)$ which
includes the derivative with respect to $x$.

If there are more interfaces as in the systems depicted in
FIG.~\ref{SNSNSfig}, multiple scattering occurs and the Green's
function is given by
\begin{eqnarray}
 G_{{\nu}j{\nu}{'}j'}(x,x',k_y,k_z,i\omega_n)&=&
 G_{{\nu}j{\nu}{'}j'}^{0} (x,x',k_y,k_z,i\omega_n)
 (\delta_{\nu\nu{'}} \delta_{jj'} + \delta_{-\nu \nu '}
 \delta_{j + \nu , j'})\nonumber\\
\label{Greenmult}
 &&+ \sum_{\sigma\mu{\sigma}{'}{\mu}{'}}d_{{\nu}j}^{\sigma}
 d_{{\nu}{'}j'}^{{\sigma}{'}}
 \psi_{{\nu}j}^{\sigma\mu}(x)T_{{\nu}j{\nu}{'}j'}^{\sigma{\sigma}{'}\mu{\mu}{'}}
 \tilde{\psi}_{{\nu}{'}j'}^{\sigma{'}\mu{'}}(x').
\end{eqnarray}
This is the generalization of Eq. (\ref{GreenNS}). There are
features in Eq. (\ref{Greenmult}) that do not appear in that
equation. One of these features is the strange combination of
Kronecker deltas in the first term. The first set of Kronecker
deltas, ${\delta}_{\nu\nu{'}}\delta_{jj'}$, insures that the
first term is nonzero only if the starting and final positions
are in the same layer. The other set,
${\delta}_{-\nu\nu{'}}{\delta}_{j+{\nu},j'}$, serves the same
purpose but at the same time it takes care of the redundancy in
the indexing of the layers. The structure of the two sets of
Kronecker deltas assures us that there is no overlapping of their
functions. If one set gives the value unity the other set must be
zero and vice versa. Another feature of Eq. (\ref{Greenmult}) is
the presence of the quantity
$T_{{\nu}j{\nu'}j'}^{\sigma{\sigma}'\mu{\mu}'}$. Whereas the
$t$-matrix describes the scattering of the particle at a
particular interface, this quantity describes the multiple
scattering of the particle at the interfaces along its path. We
call this the multiple scattering $T$-matrix. The $T$-matrix is
in fact a function of the $t$-matrix and is given by the multiple
scattering equation
\begin{equation}
\label{Tmatrix} T_{\nu j \nu ' j'}^{\sigma\sigma '\nu\mu '}=
t_{{\nu} j {\mu}{'} j'}^{{\sigma}{\sigma}{'}{\nu}{\mu}{'}}
(\delta_{{\mu}{'}{\nu}{'}} \delta_{jj'} + \delta_{-\mu'\nu'}
\delta_{j + \mu' , j'}) +\sum_{\sigma {''} \nu {''}} t_{\nu j \nu
{''} j}^{\sigma \sigma {''} \nu \nu {''}} d_{\nu {''} j}^{\sigma
{''}} T_{\nu {''} j \nu {'} j'}^{\sigma {''} \sigma {'}, - \nu
{''} \mu {'}}.
\end{equation}
To solve this equation it is necessary to first solve for the
$t$-matrix at each interface using Eq. (\ref{tmatrix}).

Until now no approximations have been made in using the solutions
(\ref{solution1}) of the Bogoliubov-de Gennes equations. In
applying the formalism given above to the calculation of the
density of states, $i\omega_n$ has to be replaced by $E+i\delta$.
After that one can make the frequently used Andreev approximation,
which amounts to the replacement
\begin{equation}
\label{AA}
k_{\nu j}^{\sigma}\to
k_{F_x}+\sigma\frac{\sqrt{E^2-|\Delta|^2}}{2k_{F_x}}
\end{equation}
if $k_{\nu j}^{\sigma}$ occurs in the exponential and to $k_{\nu
j}^{\sigma}\to k_{F_x}$ if $k_{\nu j}^{\sigma}$ occurs as a
factor as shown explicitly in the third and fourth components of
the wavefunction (\ref{solution6}). This approximation is valid
when $E,|\Delta|<<k_{F_x}^2$. In the present paper we will
investigate its limitation by looking at configurations in which
$E,\Delta\approx k_{F_x}^2$.

\section{The Local Density of States}
\label{LDOS}

Sec.~\ref{theory} provides the basic machinery to determine the
matrix Green's function, which enables us to calculate the local
density of states (LDOS) at a position $x$ using the equation
\begin{equation}
\label{LDOS1} {\rm LDOS}(x,E)= -\frac{1}{\pi L_y L_z}
\lim_{\delta \rightarrow 0} \sum_{k_y , k_z} {\rm Im} G_{11}
(x,x;k_y , k_z, E + i \delta)
\end{equation}
in which $G_{11}$ is the upper left matrix element of the
multiple scattering Green's function (\ref{Greenmult}). Although
we only study Andreev bound states, which imply infinite peaks at
the bound-state energies, the Green's function formalism makes it
possible to broaden the peaks by using a small value of $\delta$.
The peaks acquire a finite height and must correspond to the
bound-state energies.

\subsection{The SNS Junction}
\label{SNS}

We first apply the formalism discussed in Sec.~\ref{theory} to a
superconductor-normal metal-superconductor (SNS) junction. A
schematic diagram of the system studied is given in the upper and
middle panels of FIG.~\ref{SNSNSfig}. The dotted lines in the
upper panel serve as an indication of the position of the inner
NS and SN interfaces in the SNSNS system to be treated in
Sec.~\ref{SNSNS}. In this first application we will show more
explicitly how the different labels are used. The interface index
$j$ has only two values $1$ and $2$. The $T$-matrix equation
(\ref{Tmatrix}) can be recast into the form
\begin{equation}
\label{SNSTmat1} T^{- \sigma \sigma {'} - \nu \nu {'}}_{-\nu j \nu
{'} j'} =t_{-\nu j \nu {'} j'}^{\sigma \sigma {'}- \nu \nu {'}}
\delta_{jj'} + \sum_{\sigma {''} \nu {''}} t_{-\nu j \nu {''}
j'}^{\sigma \sigma {'} - \nu \nu {'}} d_{\nu {''} j}^{\sigma
{''}} T_{- \nu {''},j+ \nu {''}, \nu {'}j'}^{\sigma {''} \sigma
{'}, - \nu {''} \nu {'}}.
\end{equation}
To implement this in matrix form, we must see to it that the
elements of the $T$-matrix in both sides of the equation are
arranged in the same manner. Thus, its elements in the right-hand
side of the equation must be similarly displayed as in the
left-hand side. This can be done by writing Eq. (\ref{SNSTmat1})
in the form
\[
\left[ \begin{array}{cc} T_{- \nu 1 \nu {'}1}^{\sigma \sigma {'}
- \nu \nu {'}} &
T_{- \nu 1 \nu {'}2}^{\sigma \sigma {'} - \nu \nu {'}} \\
T_{- \nu 2 \nu {'}1}^{\sigma \sigma {'} - \nu \nu {'}} & T_{- \nu
2 \nu {'}2}^{\sigma \sigma {'} - \nu \nu {'}}
\end{array} \right]
 =\left[ \begin{array}{cc}
t_{- \nu 1 \nu {'}1}^{\sigma \sigma {'} - \nu \nu {'}} &
0 \\
0 & t_{- \nu 2 \nu {'}2}^{\sigma \sigma {'} - \nu \nu {'}}
\end{array} \right]+
\]
\begin{equation}
\label{SNSTmat2}
\left[ \begin{array}{cc} 0 &  t_{- \nu 1 \nu
{''}1}^{\sigma \sigma {''} - \nu \nu {''}}
d_{\nu {''} 1}^{\sigma {''}} \delta_{\nu {''} +} \\
t_{- \nu 2 \nu {''}2}^{\sigma \sigma {''} - \nu \nu {''}} d_{\nu
{''} 2}^{\sigma {''}} \delta_{\nu {''} -} & 0
\end{array} \right]
\left[ \begin{array}{cc} T_{- \nu 1 \nu {''}1}^{\sigma \sigma
{''} - \nu \nu {''}} &
T_{- \nu 1 \nu {''}2}^{\sigma \sigma {''} - \nu \nu {''}} \\
T_{- \nu 2 \nu {''}1}^{\sigma \sigma {''} - \nu \nu {''}} & T_{-
\nu 2 \nu {''}2}^{\sigma \sigma {''} - \nu \nu {''}}
\end{array} \right].
\end{equation}
Each element of the above matrices is itself a $4\times 4$ matrix
with $\sigma\nu$ $(=++,+-,-+,--)$ as row and column indices. So
actually, the matrices appearing in Eq. (\ref{SNSTmat2}) have
dimensions $8\times 8$. In this form the $T$-matrix can be
obtained by appropriate matrix inversion. The Green's function is
obtained using Eq. (\ref{Greenmult}). The local density of states
is determined by using Eq. (\ref{LDOS1}).

To investigate the validity of the Andreev approximation in our
SNS junction, we will focus on the choice of the transverse width
$L_y=L_z=L_t$ of the junctions. The transverse wave components of
the wavefunction (\ref{solution1}) are standing waves proportional
to $\sin(k_yy)\sin(k_zz)$ where $k_y$ and $k_z$ are given by Eq.
(\ref{transwavevector}). The different combinations of
$(k_y,k_z)$ or $(n_y,n_z)$ are called modes whose allowed values
are determined by
\begin{equation}
\label{kF} k_{F_x}^2=\mu-\left( \frac{\pi}{L_t} \right)^2
(n_y^2+n_z^2) > 0.
\end{equation}
When the transverse dimension is small, the second term in the
right becomes large, as a result, only a few modes will be
allowed. If this term exceeds the chemical potential $\mu$,
$k_{F_x}$ becomes imaginary, the wavefunction is damped, and
consequently, such mode cannot exist. For larger transverse
dimensions, the second term is smaller whereupon more modes are
allowed. Most of our calculations will be done for small $L_t$ so
that only few modes will exist. We will tune $L_t$ such that
$k_{F_x}^2$ is of the same order of magnitude as the gap energy
$\Delta$, in which regime the Andreev approximation (\ref{AA}) is
not valid, and call such a $L_t$ value a critical width.

FIGs.~\ref{PSNS2} and \ref{PSNS1} show the results for a
configuration in which $(n_y,n_z)=(2,2)$ is the highest allowed
mode. The chemical potentials in the superconductor and in the
normal metal, $\mu_S$ and $\mu_N$ respectively, are assumed equal
with magnitude 0.5. The longitudinal dimension $L$ of the
normal-metal part is $4000$ Bohr and the gap $\Delta$ is treated
as real with magnitude $0.0001$ Ry. The LDOS in the normal-metal
part at $x=1000$ Bohr is plotted against $E/\Delta$ . The peaks
represent discrete energy states \cite{Miriam}. We make the width
of the curves, determined by the the parameter $\delta$ in $E+i
\delta$, wide enough so that the fundamental features can be
seen. The numbers in parentheses denote the mode to which the
energy belongs. In FIG.~\ref{PSNS1} the transverse width is
determined by the condition that $k_{F_x}^2=\Delta$ for the mode
$(2,2)$, in which one finds that $L_t=12.5676$ Bohr and in
FIG.~\ref{PSNS2} the transverse width is $L_t=13$ Bohr, which is
slightly larger than the critical width but has the same allowed
modes. In both figures the dashed curve is the local density of
states calculated using Andreev approximation (AA) and the solid
curves are the results from the exact calculation. In
FIG.~\ref{PSNS2}, the exact and the AA results coincide and just
three states are found, one for each mode. For the critical
width, the states for the first two modes are at almost the same
position, but for the $(2,2)$ mode many states are found. We
notice that the AA peaks have about the same height whereas the
amplitudes of the exact peaks vary widely with the energy. The
amplitude variation of the exact peaks is due to the specific
position chosen for the LDOS. FIG.~\ref{PSNS3} depicts the exact
LDOS at $x=1000$ Bohr and $x=1500$ Bohr. We observe that some of
the peaks in the solid curve are suppressed while the
corresponding peaks in the dashed curve are prominent and vice
versa. The suppressed peaks are pulled down by the small
magnitude of the weighting wavefunctions at those values of $x$.
The degenerate traveling waves corresponding to the AA are split
in the exact treatment into odd and even (sine and cosine)
functions having different heights at different positions
\cite{Gyorffy}.

Finally, we want to comment on our choice of the transverse width
up to now. In Sec.~\ref{gap} it will come out that
superconductivity is suppressed for transverse widths in the
order of $20$ Bohr or less. This means that so far our choice of
transverse widths may seem not too appropriate. We chose those
transverse widths to illustrate with clarity the fundamental
features of the bound states. In order to see the influence of a
wider transverse dimension, we show in FIG.~\ref{PSNS4} the LDOS
of a SNS system with $L_t=100$ Bohr (solid curve) and
$L_t=99.8514$ Bohr (dotted curve). The critical transverse width
of $99.8514$ Bohr is obtained from the condition
$k_{F_x}^2=\Delta$ for the mode $(19,12)$. To illustrate its main
features clearly, we only show the results for the Andreev
approximation. It appears that higher modes are allowed for a
wider transverse dimension. The peak at about $E=0.95\Delta$ comes
from the many lower modes, each supporting just one bound state.
Only the highest mode gives rise to the distinct set of peaks
starting at about $E=0.074\Delta$. So effectively, the character
of the results shown above is essentially unchanged.

\subsection{The SNSNS System}
\label{SNSNS}

A schematic representation of the system is shown in the lower
panel of FIG.~\ref{SNSNSfig}. The height of the middle
superconductor is chosen to take the values $h \Delta$ where the
range of $h$ is $0 \leq h \leq 1$. The LDOS is calculated using
Eq. (\ref{LDOS1}). The process of determining the $t$-matrix, the
$T$-matrix, and the Green's function for the SNSNS junction is
the same as in the SNS junction. The number of interfaces has
increased by $2$, and the dimension of the matrices has become
$16$ instead of $8$. We again choose a square transverse cross
section whose dimension is $L_t=12.5676$ Bohr. This means that the
set of modes is the same as in the SNS system. The lengths of the
normal metal parts are each $1000$ Bohr and that of the middle
superconductor is $2000$ Bohr.

FIG.~\ref{PSNSNS} shows the development of the bound states as
the gap of the middle superconductor is increased from
$0.25\Delta$ to $\Delta$. All peaks belong to the highest $(2,2)$
mode, apart from a $(1,1)$ mode and a $(1,2)$ mode peaks just
below $E/\Delta=1$. The mode labels are shown only in
FIG.~\ref{PSNSNS}d to avoid cluttering the figures. Both the exact
solution and the states according to the Andreev approximation
(AA) are shown. Again the lifting of degeneracy is observed in
the exact results. The distribution of the peak heights shown
corresponds to a calculation of the LDOS at $x=-1500$ Bohr.
Results for other positions just give other distributions of peak
heights. From now on, we concentrate on the position of the AA
peaks to facilitate the comparison of the different figures. The
SNSNS system with $h=0$ is equivalent to the SNS system, so it is
interesting to compare FIG.~\ref{PSNS1} with
FIGs.~\ref{PSNSNS}a-d. One can see a general shift of the states
in the latter figure to the right relative to the states in
FIG.~\ref{PSNS1}. For the system with $h=0.25$ the $h=0$ state at
$E=0.075\Delta$ suffers a large displacement. This can be
understood as follows. States below $E=0.25\Delta$ see a
longitudinal length $L_N$ of $1000$ Bohr whereas the states above
$E=0.25\Delta$ see a length $L_N$ of $4000$ Bohr, $L_N$ being the
length of the N metal. According to a semiclassical result
\cite{Nazarov,Ashida}, bound-state energies scale as
$\frac{v_F}{L_N}$, so they are inversely proportional to the
longitudinal length, which would imply a shift upwards to
$E=0.3\Delta$. However, the separation of $2000$ Bohr between the
N part is smaller than the BCS coherence length
$\frac{v_F}{\pi\Delta}\approx 4500$ Bohr,  which suggests that
the N parts of the system are not completely decoupled yet. This
is the reason why the lowest state is found below $E=0.25\Delta$,
namely at $E=0.19\Delta$. The positions of the other peaks are
hardly changed.

Looking at FIGs.~\ref{PSNSNS}b-d one sees that the lowest state
stabilizes at a position of about $E=0.26\Delta$. All other
states are shifted more and more upwards for increasing $h$
values. For $h=0.5$, FIG.~\ref{PSNSNS}b, still a set of peaks is
found above $0.5\Delta$, in FIG.~\ref{PSNSNS}c the set starts at
about $E=0.75\Delta$. For $h=1$, FIG.~\ref{PSNSNS}d, only three
states are seen just below $E=\Delta$, apart from a state at
$E=0.78\Delta$, lying three times as high as the lowest state, in
agreement with the semiclassical picture. A test calculation for
a SNS system with $L_N=1000$ Bohr differs from FIG.~\ref{PSNSNS}d
only as far as the position of the highest peak, the $(1,1)$ peak,
is concerned.

The idea of decoupling can also be illustrated by
FIG.~\ref{Pcoupling}. For the sake of clarity we just give the
results for the Andreev approximation. FIG.~\ref{Pcoupling}(a)
shows the LDOS for a SNSNS system with $L=6000$ Bohr. The lengths
of the N parts are kept at $1000$ Bohr but now the length of the
middle superconductor is $4000$ Bohr, slightly below the BCS
coherence length of approximately $4500$ Bohr. In
FIG.~\ref{Pcoupling}(b), the length $L$ of the SNSNS system is
$8000$ Bohr and the lengths of the N parts are again kept at
$1000$ Bohr. This makes the middle superconductor $6000$ Bohr in
length, which is longer than the BCS coherence length. In both
figures the gap of the middle superconductor is $0.25\Delta$.
Thus, there is effectively complete decoupling in
FIG~\ref{Pcoupling}(b). In addition, one sees that the $(2,2)$
peaks are lying closer to one another for the longer system,
which is in line with the $\frac{V_F}{L_N}$ behavior of the
energies.

\section{Self-Consistent Calculation of the Gap}
\label{gap}

In the preceding discussions, the gap function $\Delta$ used in
the calculation is steplike, that is, it has a finite constant
value in the superconducting part and is zero in the normal part.
At the interfaces, it has a step discontinuity. This
configuration is shown in FIG.~\ref{SNSNSfig}. So in the
calculation of the bound states in Sec.~\ref{LDOS}, the proximity
effect is not taken into account. However we want to demonstrate
in this section that our formalism makes it possible to show that
the actual gap function is not steplike, but decreases smoothly
towards the interface and abruptly goes down to zero in the
normal metallic layer. This proximity effect is studied in the
present section. In addition, the actual gap function will be
calculated self-consistently.

The self-consistency condition is given by
\begin{equation}
\label{selfcon1} \Delta(x)=-VF({\bf r,r},0^+)
\end{equation}
in which $F({\bf r,r},0^+)$ is the well-known anomalous Green's
function  \cite{Gorkov} given by the upper right element of the
original $(\bf {r},\tau)$-dependent matrix Green's function.
Taking into account the expansion of the matrix Green's function
over the Matsubara frequencies and the transverse wave vectors,
we obtain
\begin{equation}
\label{selfcon2} \Delta(x)=-\frac{V}{\beta L_y
L_z}\sum\limits_{\omega_n,k_y,k_z} F(x,x,k_y,k_z,i\omega_n)
\end{equation}
where $F(x,x,k_y,k_z,i\omega_n)$ is the upper right element of
the matrix Green's function in Eq. (\ref{Greenmult}). The
summation over the Matsubara frequencies is restricted by the
Debye temperature $\theta_D$ according to the formula,
\begin{equation}
\label{DebyeT}
 k_B\theta_D=\omega_{n_{max}}=n_{max}\pi k_B T.
\end{equation}
The summation over the transverse wave vector is over all positive
values of $k_y$ and $k_z$ according to Eq.
(\ref{transwavevector}). Evaluation of the summation takes much
computer time, so we resort to some approximations. Since the
cross section is a square, that is $L_y=L_z$, an excellent
approximation to reduce the number of terms is to partition the
transverse $k_\perp$-plane by concentric circles with radius
$k_\perp = \sqrt{k_y^2+k_z^2}$ and a $\delta k_\perp
=\frac{\pi}{L_t}$. The number of allowed values of $k_y$ and
$k_z$ in each ring is given by its surface divided by the density
of the transverse $k$ states $\left(\frac{\pi}{L_t}\right)^2$.
Another approximation can be made by noting that the terms
involving $k_\perp
>>\sqrt{\mu}$ do not have significant contributions to the sum.
This allows us to evaluate a finite number of terms instead of
evaluating an infinite series. In our calculation we take
$k_{\perp_{max}}=3\sqrt{\mu}$. In determining the coupling
constant $V$ according to the BCS-relation
\begin{equation}
\label{Tc} T_c=1.13~\omega_D~e^{\frac{-1}{N(\mu)V}}, ~{\rm
with}~N(\mu)=\frac{\mu}{4\pi ^2}
\end{equation}
we use $T_c=1.2$ K and $\omega_D=375$ K for aluminum. We find
$V=9.516$ Ry.

\subsection{The Bar-Shaped Superconductor}
\label{gapS}

The matrix Green's function for a bar-shaped superconductor
$G_S^0 (x,x,k_y,k_z,i\omega_n)$ is given by Eq. (\ref{Green3}).
By substituting the upper right element of this Green's function
in Eq. (\ref{selfcon2}) one straightforwardly obtains the
following selfconsistency condition for the homogeneous
bar-shaped superconductor,
\begin{equation}
\label{selfcon3} \Delta(x)=\frac{V|\Delta|}{\beta L_y L_z}
 \sum \limits_{\omega_n,k_y,k_z}
 \frac{1}{4\Omega_S}
 \left(\frac{1}{\sqrt{k_{F_x}^2 + i\Omega_S}}
     + \frac{1}{\sqrt{k_{F_x}^2 - i\Omega_S}}
 \right).
\end{equation}

In calculating the gap self-consistently,  we first assume an
initial value of the gap. By substituting it on the right-hand
side of Eq. (\ref{selfcon3}) we obtain a new value of the gap.
Then this new value is again substituted on the right-hand side
to get another new value. This procedure can be repeated until the
difference between successive iterations is negligibly small. As
shown in FIG.~\ref{PGapite}, the difference between the first ten
iterations is still significant. After about 80 iterations, the
value of the gap stabilizes to $1.106\times 10^{-5}$ Ry. The
transverse length is $1000$ Bohr and the temperature is $0.6$ K.
The initial value of the gap is $2.0 \times 10^{-5}$ Ry.

FIG.~\ref{PGapSLt} shows the plot of the gap against the
transverse length $L_t$ for different temperatures. The number of
iterations is $100$. It can be seen that there are oscillations
of the gap. The amplitudes of the oscillations decrease as the
transverse dimension increases. These oscillations can be
attributed to the discreteness of the transverse wave vector. As
the transverse width increases, the transverse wave vector
approaches the continuous regime which can be gauged from the gap
becoming closer to its bulk value obtained by integrating instead
of summing over the transverse wave vectors. In the figure, we
show the bulk value at $0.2$ K. Another interesting thing which
can be seen in the figure is the suppression of superconductivity
for narrower transverse dimensions. We notice that as the
temperature increases the onset of the suppression of
superconductivity occurs at higher values of $L_t$.

In FIG.~\ref{PGapST} we show the temperature variation of the gap
for different transverse widths $L_t$. A residual value of the
gap for the bulk superconductor can still be observed beyond the
critical temperature ($1.2$ K for aluminum). We also notice
small-amplitude oscillations of the gap at higher temperatures,
which only show up in the curves for larger transverse
dimensions. For smaller transverse dimensions, the oscillations
are suppressed. These oscillations are due to the cut-off in the
summation over the Matsubara frequencies. For lower temperatures,
the cut-off value $n_{max}$ in Eq. (\ref{DebyeT}) becomes very
large and the results are no longer sensitive to it.

\subsection{The NS System}
\label{gapNS}

The matrix Green's function appropriate for a normal
metal-superconductor system is given in Eqs. (\ref{GreenNS}) and
(\ref{auxGreenNS}). The first term on the right-hand side of Eq.
(\ref{GreenNS}) is the matrix Green's function for the bar
superconductor. The second term contains the elements of the
$t$-matrix, which take into account the scattering of the
quasiparticles at the interface. The latter acts as a perturbing
term to the former and is therefore responsible for the spatial
variations of the gap and the pair amplitude at the vicinity of
the interface. To calculate the spatial variation of the gap
according to Eq. (\ref{selfcon2}), we substitute the upper right
element of the matrix Green's function, Eq. (\ref{GreenNS}),
using the value of the self-consistent gap for the bar-shaped
superconductor.

Before presenting our results, we want to mention a computational
problem which has to be solved. Due to the explicit presence of
the position $x_j$ in Eq. (\ref{tmatrix}), the matrices involved
have alternating columns of very small and very large values
which the computer cannot handle anymore. However this problem is
not intrinsic to the formalism and can be solved by appropriate
rescaling. By defining the matrix $\hat{t}$
\begin{equation}
\label{t-transform1} \hat{t}_{\nu j\nu 'j}^{\sigma \sigma '\nu \nu
'} \equiv e^{i\sigma \nu k_{\nu j}^\sigma x_j} t_{\nu j\nu
'j}^{\sigma \sigma '\nu \nu '} e^{i\sigma ' \nu 'k_{\nu
'j}^{\sigma '} x_j}
\end{equation}
the Green's function (\ref{GreenNS}), combined with Eq.
(\ref{auxGreenNS}), obtains the form
\begin{eqnarray}
 G_{\nu j\nu 'j}(x,x',k_y,k_z,i\omega_n)&=&
 \sum_{\sigma}d_{\nu j}^{\sigma}
 \psi_{\nu j}^{\sigma\mu}(x)
 \tilde{\psi}_{\nu j}^{\sigma ,-\mu }(x')
 \delta_{\nu \nu '}\nonumber\\
 \label{GreenNS1}
 & &+ \sum_{\sigma\sigma '}d_{\nu j}^{\sigma}
 d_{\nu 'j'}^{\sigma '}
 \psi_{\nu j}^{\sigma\nu}(x-x_j)
 \hat{t}_{\nu j \nu 'j}^{\sigma \sigma ' \nu \nu '}
 \tilde{\psi}_{\nu 'j}^{\sigma ' \nu '}(x'-x_j)
\end{eqnarray}
in which only position differences occur. The rescaled matrix
$\hat{t}$ is determined by the equation
\begin{equation}
\label{t-transform2} \sum\limits_{\sigma\nu} \nu d_{\nu j}^\sigma
\psi_{\nu j}^{\sigma\nu}(0)
 \hat{t}_{\nu j\nu 'j}^{\sigma \sigma '\nu \nu '}
 = -\nu '\psi_{\nu 'j}^{\sigma '\nu '}(0),
\end{equation}
found straightforwardly from the original equation
(\ref{tmatrix}).

FIG.~\ref{PGapNS}(a) shows the spatial variation of the gap near
the interface of a normal metal-superconductor system at $T=0.6$ K
for different transverse widths. At a distance of 30000 Bohr from
the interface, which is about six times the coherence length
($\approx 4500$ Bohr), the gap is slightly smaller than the one
obtained for the bar-shaped  superconductor. The differences are
about $7.79\%$, $6.57\%$, and $2.93\%$ of their bar value for
$L_t=99.8514$, $L_t=100$,
 and $1000$ Bohr, respectively. At $4000$ Bohr, which is of the order of the
coherence length from the interface, the differences are about
$21.7\%$, $20.44\%$, and $14.71\%$, respectively. These figures
lead to the inevitable conclusion that for larger transverse
dimensions, the proximity effect is less pronounced than for
smaller ones. This may be due to the fact that for small
transverse dimensions the superconductivity tends to be
suppressed. We find that the difference seen for the two smaller
widths can be attributed to the oscillations in the
self-consistent gap shown in FIG.~\ref{PGapSLt}. We have not seen
a special influence of the fact that the smaller width is a
critical one. FIG.~\ref{PGapNS}(b) shows the corresponding pair
amplitude defined by $\frac{\Delta (x)}{V}$ (see Eq.
(\ref{selfcon1})) which has a finite value in the normal metallic
region near the NS interface but it decays in the inner region of
the normal metal.

\subsection{The SNS System}
\label{gapSNS}

In the superconductor-normal metal-superconductor system, there
are two interfaces which we designate as $x_1$ and $x_2$. The
$t$-matrices must be determined at these interfaces so that we can
evaluate the $T$-matrix. The unpleasant singularities due to the
explicit presence of the position of the interfaces in Eq.
(\ref{tmatrix}) can be removed by using Eq. (\ref{t-transform1})
resulting to a corresponding transformation of the $T$-matrix
given by
\begin{equation}
\label{T-transform} \hat{T}_{\nu j\nu 'j'}^{\sigma \sigma '\nu
\nu '} =e^{i\sigma \nu k_{\nu j}^\sigma x_j} T_{\nu j\nu
'j'}^{\sigma \sigma '\nu \nu '} e^{i\sigma ' \nu 'k_{\nu
'j'}^{\sigma '} x_{j'}}.
\end{equation}
By implementing this, the multiple-scattering Green's function
(\ref{Greenmult}) and Eq. (\ref{Tmatrix}), which determines the
$T$-matrix in terms of the $t$-matrices, can be modified
straightforwardly.

The steps in calculating the gap and the pair amplitude
self-consistently are the same as in Sec.~\ref{gapNS}.
FIG.~\ref{PGapSNS1} shows the self-consistent gap function and the
pair amplitude for different transverse widths. The center of the
system is at $x=0$. The spatial variation of the gap near the
interfaces is clearly shown. The proximity effect is stronger
than for the NS system shown in FIG.~\ref{PGapNS}. Whereas in
FIG.~\ref{PGapNS}a for $L_t=1000$ Bohr the gap at a distance of
$6000$ Bohr from the interface has a value of $9.753\times
10^{-6}$ Ry, in FIG.~\ref{PGapSNS1}a it has already increased to
$1.074\times 10^{-5}$ Ry, which is much closer to the bulk value
of $1.106\times 10^{-5}$ Ry. In FIG.~\ref{PGapSNS1}b the pair
amplitude in the N region does not decrease below a value of
$5\times 10^{-7}$. In FIG.~\ref{PGapNS}b it decreases to zero and
the value at $2000$ Bohr from the NS interface has already
decreased to $2.784\times 10^{-7}$. In FIG.~\ref{PGapSNS2} we show
the gap functions for different values of $L$. It can be seen
that for larger $L$ the gap function is lesser in magnitude. This
is another manifestation of the proximity effect.

\subsection{The SNSNS System}
\label{gapSNSNS}

The extension of the steps outlined in Sec.~\ref{gapSNS} leading
to the $T$-matrix and the matrix Green's function for a SNSNS
system is straightforward. In this case we have to consider four
interfaces but the procedure is basically the same. In the outer
superconductors, we again use the self-consistent value of the
gap for the superconducting bar we have calculated in
Sec.~\ref{gapS}. In calculating the gap in the middle
superconductor self-consistently, we apply the recipe introduced
by Tanaka and Tsukada \cite{Tanaka}. We start the iteration
procedure by taking half of the gap in the outer superconductors
for the gap in the middle. By substituting these values in the
righthand side of Eq. (\ref{selfcon2}), the middle $x$-dependent
gap function is obtained and its spatial average is determined.
This average value is used as the new value of the gap in the
middle superconductor. The process is repeated until the
difference of these average values between successive iterations
is negligibly small. To determine the gap's spatial variation, the
self-consistent average value of the gap in the middle
superconductor and the bar value in the outer ones are
substituted in the right-hand side of Eq. (\ref{selfcon2}).

FIG. \ref{PGapSNSNS1} shows the gap and the pair amplitude of two
SNSNS systems, one with a length $L_N=3000$ Bohr for the normal
metallic parts and the other one with a larger length $L_N=5000$
Bohr. First note that the gap of the middle superconductor is
smaller than the gap of the outer superconductors although the
metallic parameters for these superconductors  are taken equal.
This supports our choice of a $h$ value smaller than $1$ in
Sec.~\ref{LDOS}, see FIG.~\ref{SNSNSfig}. Further note that the
gap and the pair amplitude are higher for the system with narrower
normal-metal part, which is just another manifestation of the
proximity effect. Although by definition the gap is zero in the
normal-metal parts, the pair amplitude is different from zero.
This is also a manifestation of the proximity effect. If the
width of the normal-metal layers had been much larger than the sum
of the coherence lengths of the outer and middle superconductors,
the pair amplitude would have been zero there except in those
regions near the interfaces.

\section{Conclusions and Future Prospects}
\label{Conclusion}

The multiple-scattering Green's function formalism developed by
Koperdraad \cite{Koperdraad} has been applied to determine the
Andreev bound states in SNS and SNSNS junctions through the local
density of states and to calculate the superconducting gap
function self-consistently. We have shown that for transverse
junction widths tuned such that the motion in the longitudinal
direction is very slow, the Andreev approximation breaks down.
For these transverse dimensions, the highest mode is supported by
many bound states whose degeneracy is lifted when no
approximation is applied. The thickness used for the normal
metallic layers is chosen such that the lower modes are each
supported by only one nondegenerate state. Results for the
self-consistent gap functions exhibit various features of the
proximity effect. Furthermore, our results show that for small
transverse dimensions, superconductivity is suppressed.

The formalism is applicable to systems of an arbitrary number of
layers. In addition, it allows for the calculation of the
supercurrents through such junctions, to which self-consistent
gap functions are necessary \cite{Tanaka,Koperdraad}. This will be
investigated in the near future.

\newpage

\section*{Caption for figures}

\begin{figure}[ht]
\centerline{\epsfig{figure=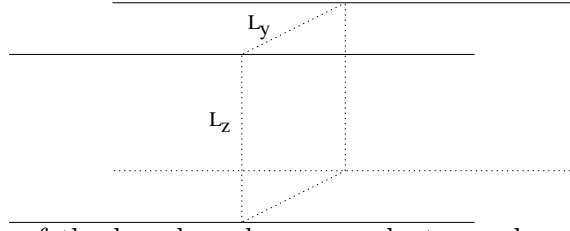,height=3.0cm}} \caption[]{The
geometry of the bar-shaped superconductor under consideration. It
has finite transverse dimension and is infinite longitudinally. A
rectangular cross-section is shown whose dimensions are $L_y$ and
$L_z$.} \label{bar}
\end{figure}

FIG.~\ref{bar}. The geometry of the bar-shaped superconductor
under consideration. It has finite transverse dimension and is
infinite longitudinally. A rectangular cross-section is shown
whose dimensions are $L_y$ and $L_z$.

\vspace{7mm}

\begin{figure}[ht]
\centerline{\epsfig{figure=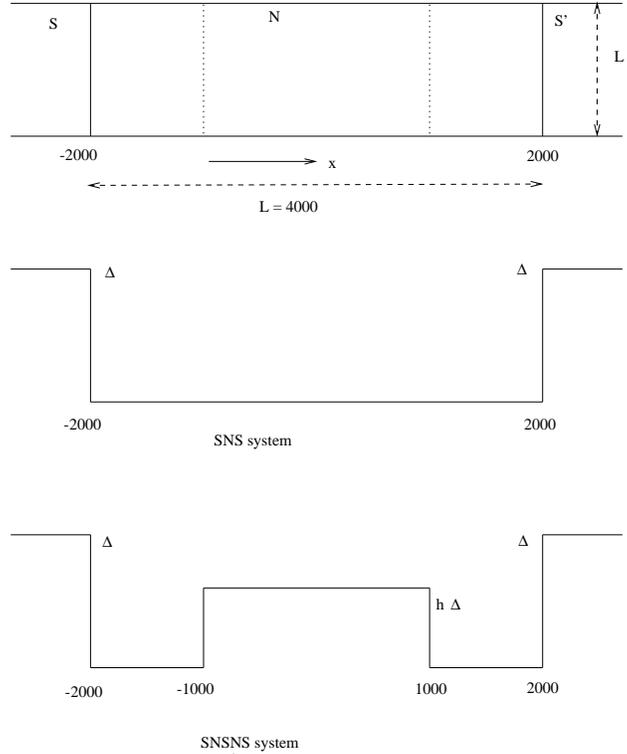,height=10.0cm}} \caption[]{A
two-dimensional picture of the systems studied. In the upper
panel the vertical direction stands for a transverse direction. In
the lower panels the potential is shown, which is zero in the
normal part(s), and proportional to $\Delta$ in the
superconducting parts.} \label{SNSNSfig}
\end{figure}

FIG.~\ref{SNSNSfig}. A two-dimensional picture of the systems
studied. In the upper panel the vertical direction stands for a
transverse direction. In the lower panels the potential is shown,
which is zero in the normal part(s), and proportional to $\Delta$
in the superconducting parts.

\vspace{7mm}

\begin{figure}[ht]
\centerline{\epsfig{figure=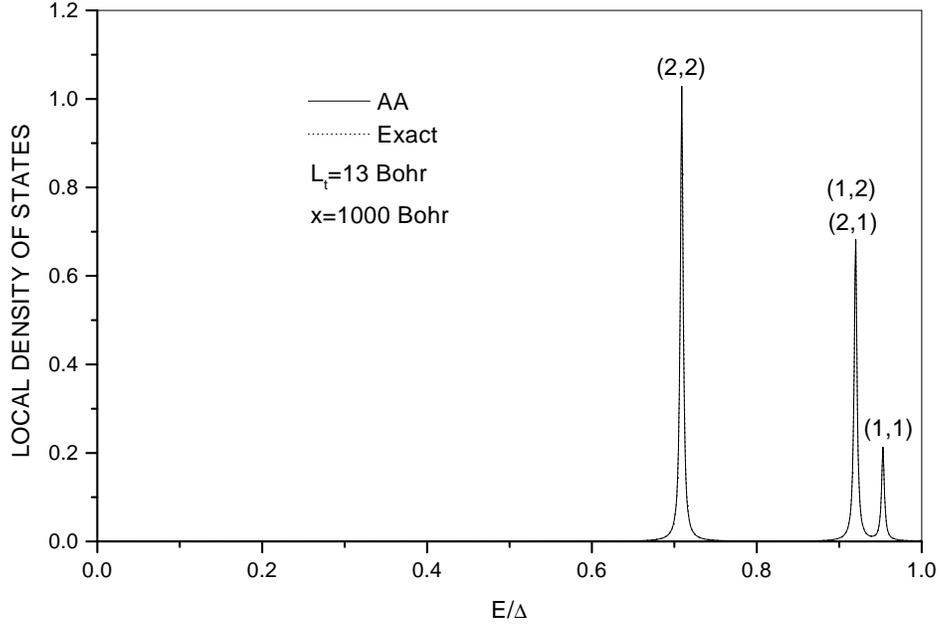,height=10.0cm}}
\caption[]{Plot of the LDOS against $E/\Delta$ for a SNS system in
which $L_t=13$ Bohr.} \label{PSNS2}
\end{figure}

FIG.~\ref{PSNS2}. Plot of the LDOS against $E/\Delta$ for a SNS
system in which $L_t=13$ Bohr.

\vspace{7mm}

\begin{figure}[ht]
\centerline{\epsfig{figure=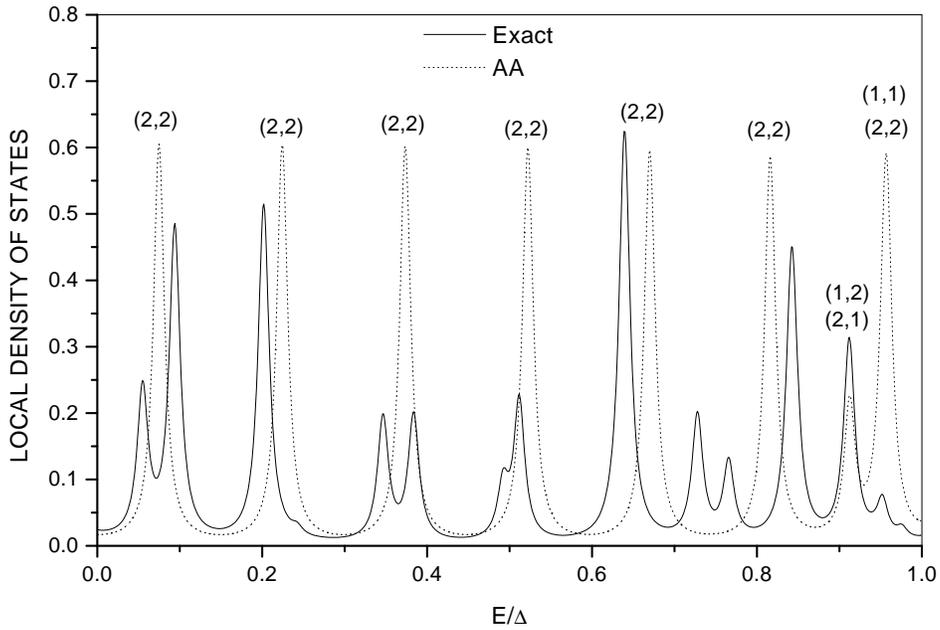,height=10.0cm}}
\caption[]{Plot of the LDOS against $E/\Delta$ for $x=1000$ Bohr,
$L_t=12.5676$ Bohr, L=4000 Bohr, and $\Delta=0.0001$ Ry.}
\label{PSNS1}
\end{figure}

FIG.~\ref{PSNS1}. Plot of the LDOS against $E/\Delta$ for $x=1000$
Bohr, $L_t=12.5676$ Bohr, L=4000 Bohr, and $\Delta=0.0001$ Ry.

\vspace{7mm}

\begin{figure}[ht]
\centerline{\epsfig{figure=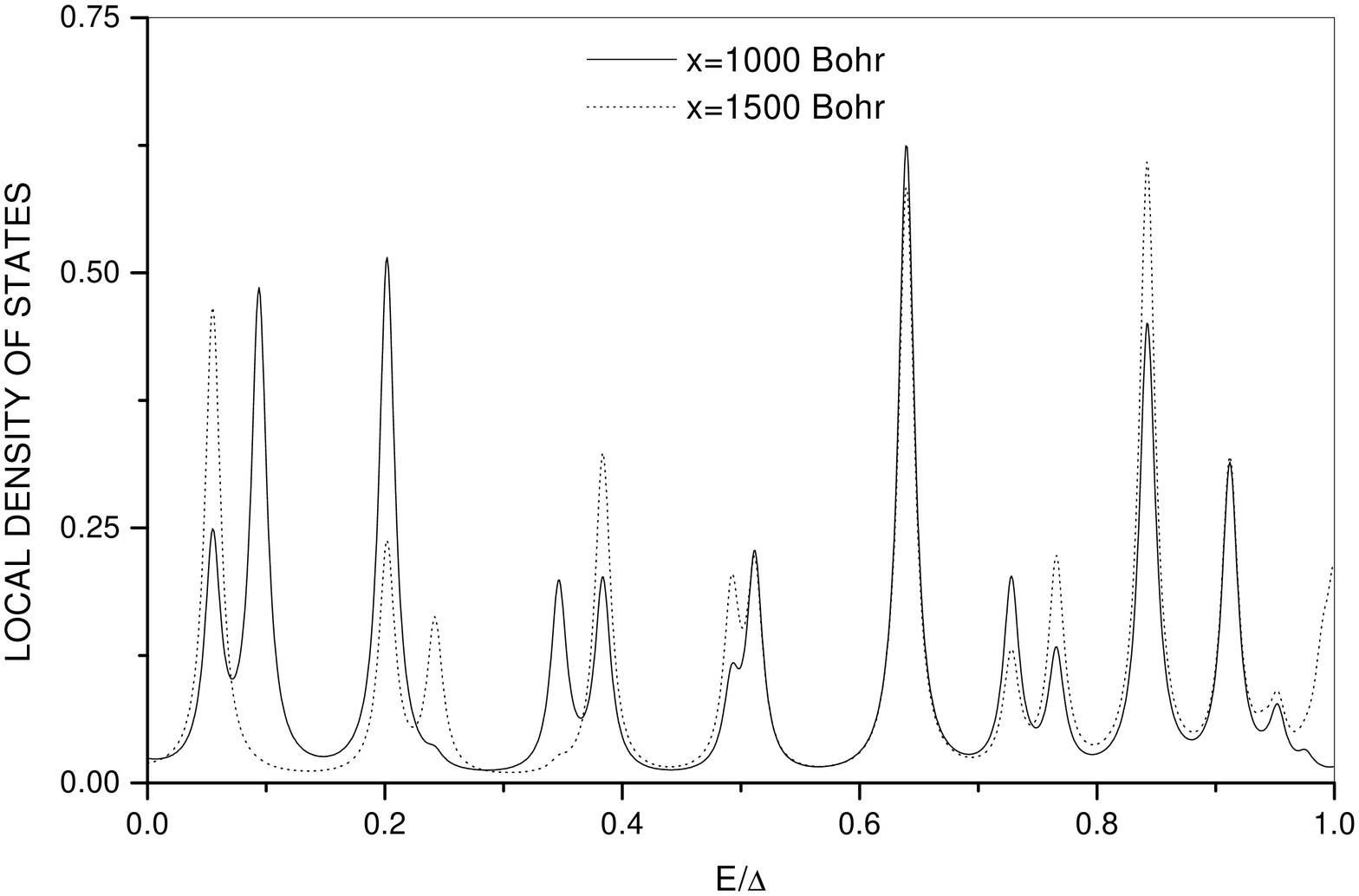,height=10.0cm}}
\caption[]{Plot of the LDOS for a SNS system against $E/\Delta$
at $x=1000$ Bohr (solid curve) and $x=1500$ Bohr (dashed curve).
The transverse dimension is $L_t=12.5676$ Bohr and the length of
the normal metal is $L=4000$ Bohr.} \label{PSNS3}
\end{figure}

FIG.~\ref{PSNS3}. Plot of the LDOS for a SNS system against
$E/\Delta$ at $x=1000$ Bohr (solid curve) and $x=1500$ Bohr
(dashed curve). The transverse dimension is $L_t=12.5676$ Bohr
and the length of the normal metal is $L=4000$ Bohr.

\vspace{7mm}

\begin{figure}[ht]
\centerline{\epsfig{figure=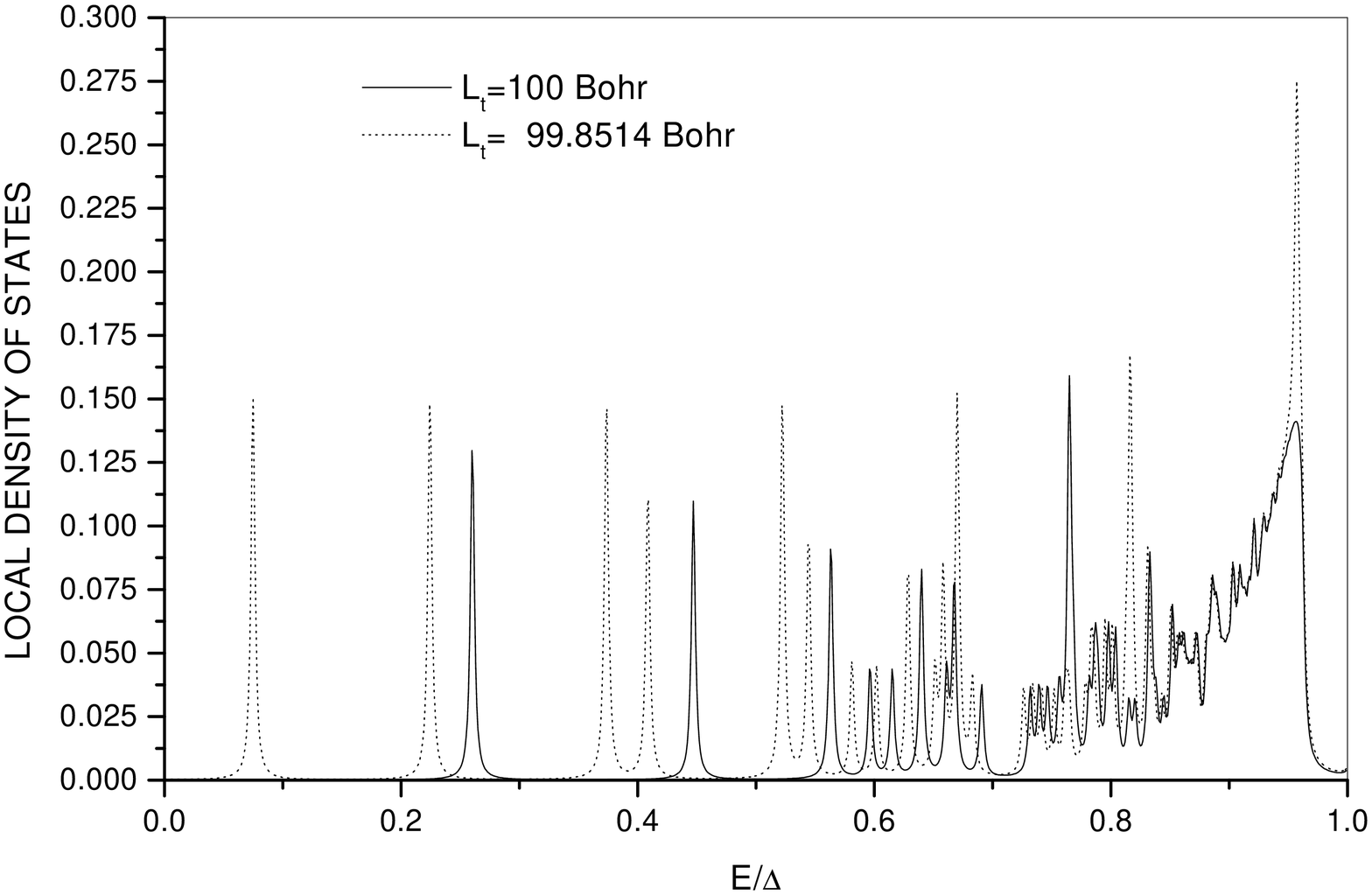,height=10.0cm}}
\caption[]{Plot of the LDOS against $E/\Delta$ for a SNS
 system in which $L_t=100$ Bohr (dotted curve)  and $L_t=99.8514$
Bohr (solid curve). The length of the normal-metal part is $4000$
Bohr.} \label{PSNS4}
\end{figure}

FIG.~\ref{PSNS4}. Plot of the LDOS against $E/\Delta$ for a SNS
 system in which $L_t=100$ Bohr (solid curve)  and $L_t=99.8514$
Bohr (solid curve). The length of the normal-metal part is $4000$
Bohr.

\vspace{7mm}

\begin{figure}[ht]
\setlength{\unitlength}{1cm}
\begin{picture}(15,15)
\put(1,1){\shortstack[c]{\epsfig{figure=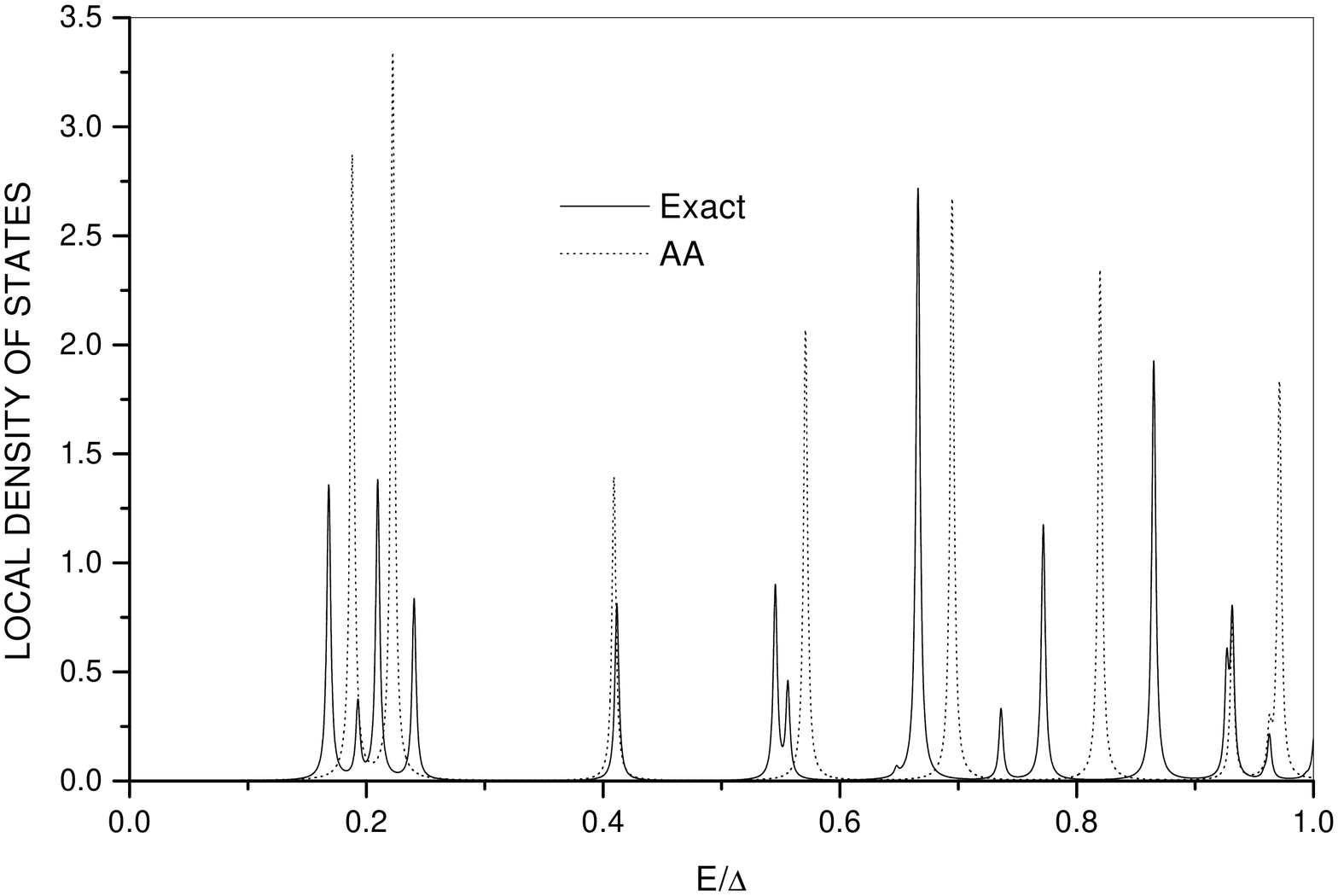,height=6.0cm}\\
\vspace{0.3cm}\epsfig{figure=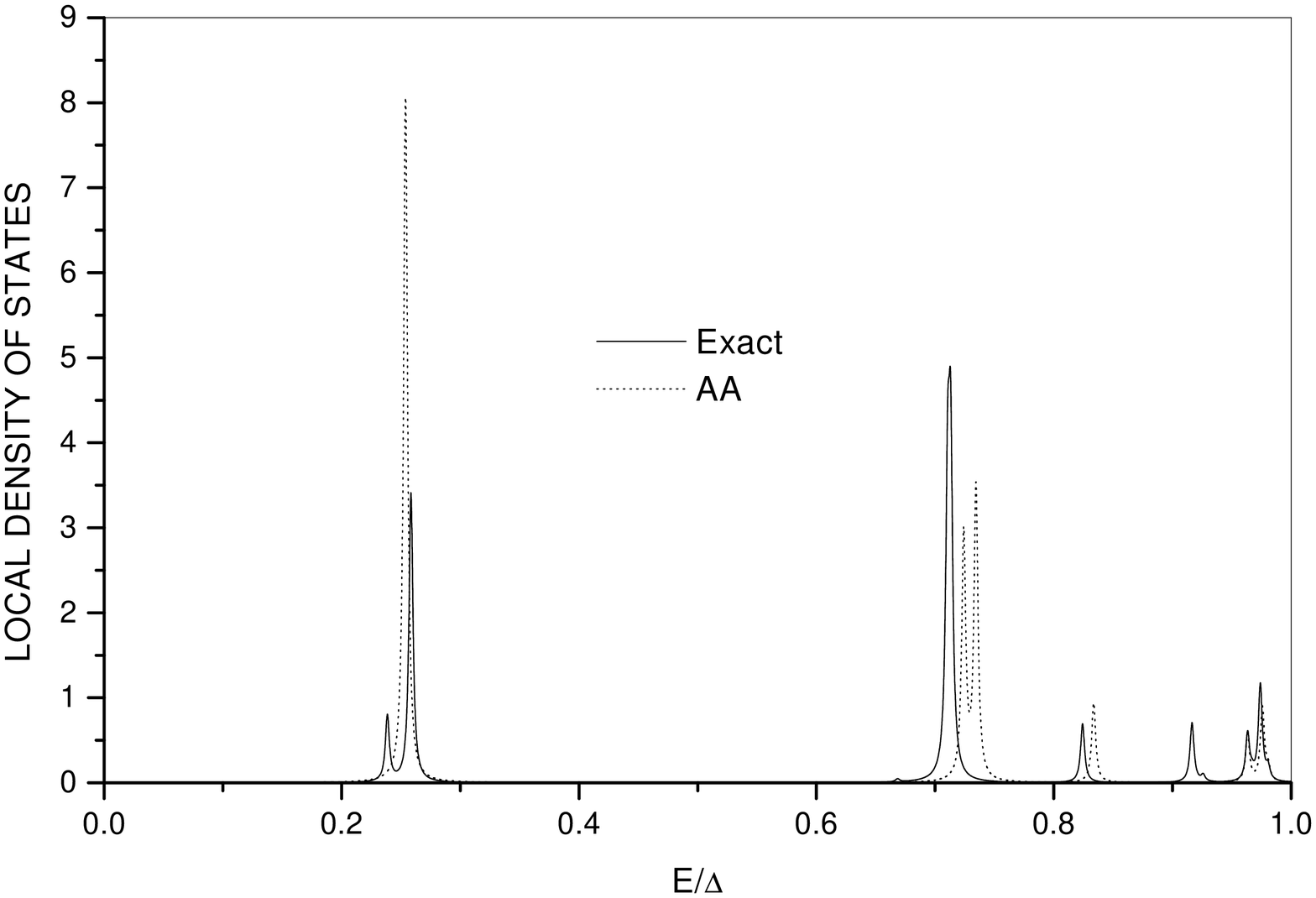,height=6.0cm}}}
\put(9,1){\shortstack[c]{\epsfig{figure=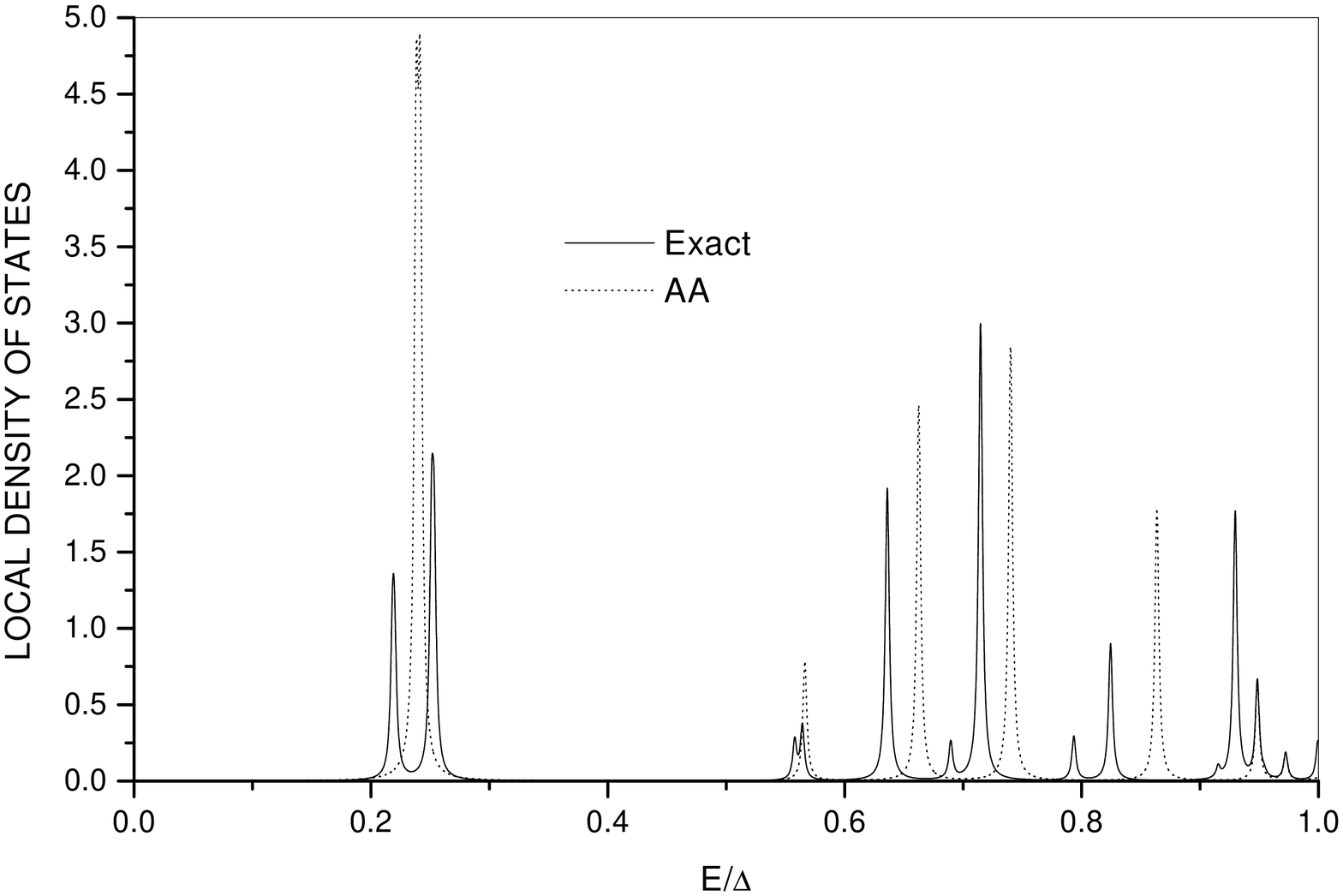,height=6.0cm}\\
\vspace{0.3cm}\epsfig{figure=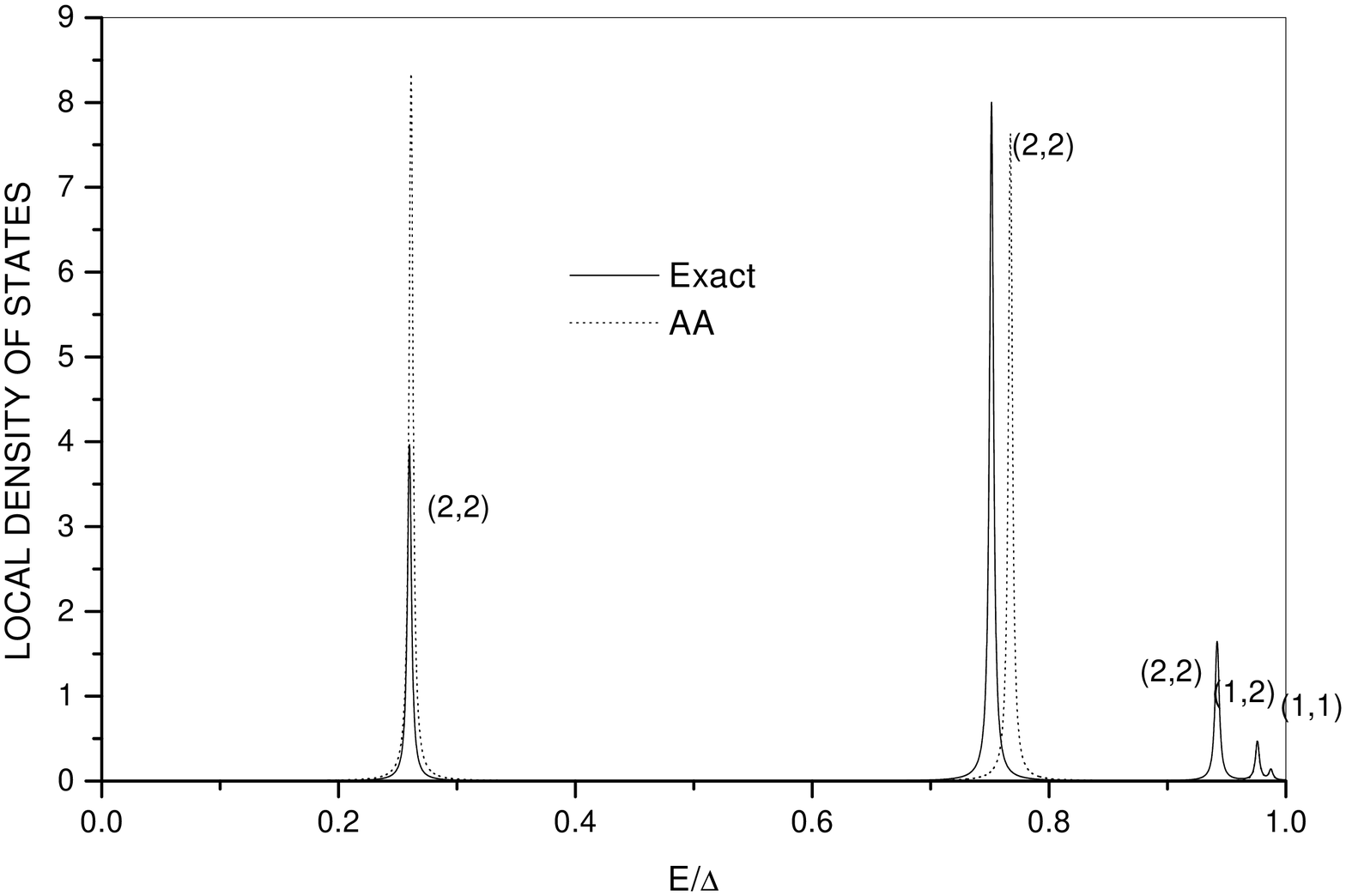,height=6.0cm}}}
\put(5.2,7.2){(a)}
\put(13.2,7.2){(b)}
\put(5.2,1){(c)}
\put(13.2,1){(d)}
\end{picture}
\caption[]{Plot of the LDOS at $x=-1500$ Bohr against $E/\Delta$
for a SNSNS system in which $L_t=12.5676$ Bohr for (a) $h=0.25$
(b) $h=0.5$ (c) $h=0.75$ and (d) $h=1$.} \label{PSNSNS}
\end{figure}

FIG.~\ref{PSNSNS}. Plot of the LDOS at $x=-1500$ Bohr against
$E/\Delta$ for a SNSNS system in which $L_t=12.5676$ Bohr for (a)
$h=0.25$ (b) $h=0.5$ (c) $h=0.75$ and (d) $h=1$.

\vspace{7mm}

\begin{figure}[ht]
\begin{center}
\setlength{\unitlength}{1cm}
\begin{picture}(18,18)
\put(3.5,1){\shortstack[c]{\epsfig{figure=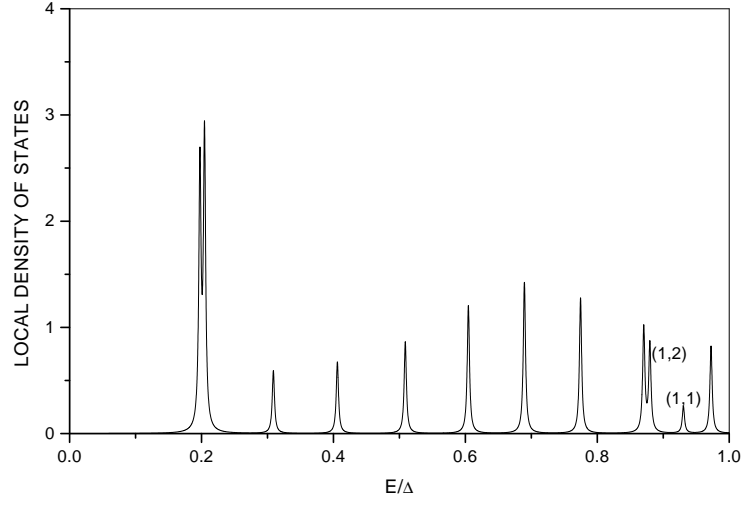,height=8.0cm}\\
\epsfig{figure=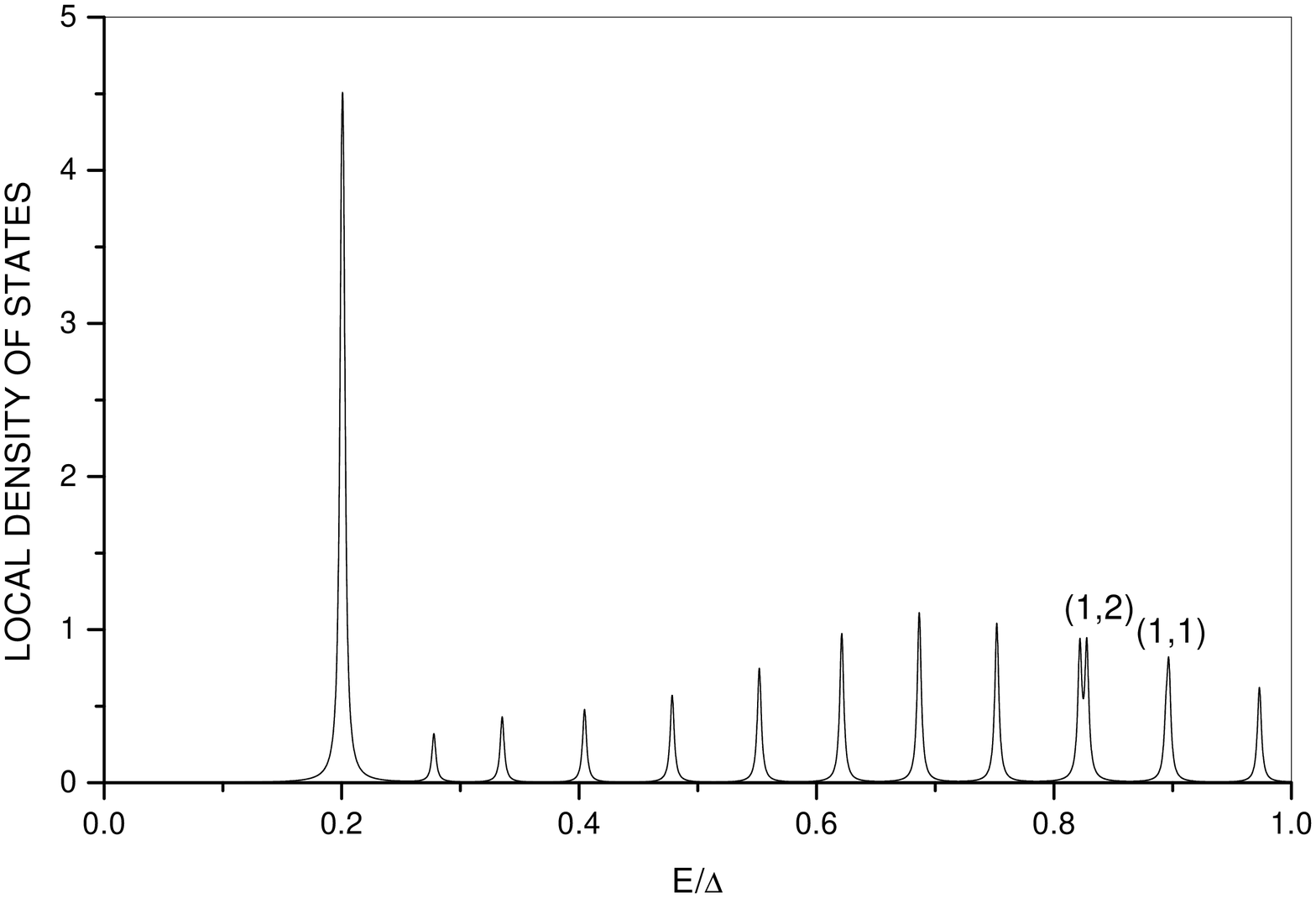,height=8.0cm}}} \put(9.3,8.8){(a)}
\put(9.3,0.5){(b)}
\end{picture}
\end{center}
\caption[]{The LDOS against $E/\Delta$ for a SNSNS system in
which $L_t=12.5676$ Bohr. In (a) $L=6000$ Bohr, the length of the
middle superconductor is $4000$ Bohr and the LDOS is calculated
at $x=-2500$ Bohr. In (b) $L=8000$ Bohr, the length of the middle
superconductor is $6000$ Bohr and the LDOS is calculated at
$x=-3500$ Bohr. The gap of the middle superconductor is
$0.25\Delta$. All unlabeled LDOS peaks belong to the mode
$(2,2)$. The calculation is done in the Andreev approximation.}
\label{Pcoupling}
\end{figure}

FIG.~\ref{Pcoupling}. The LDOS against $E/\Delta$ for a SNSNS
system in which $L_t=12.5676$ Bohr. In (a) $L=6000$ Bohr, the
length of the middle superconductor is $4000$ Bohr and the LDOS
is calculated at $x=-2500$ Bohr. In (b) $L=8000$ Bohr, the length
of the middle superconductor is $6000$ Bohr and the LDOS is
calculated at $x=-3500$ Bohr. The gap of the middle
superconductor is $0.25\Delta$. All unlabeled LDOS peaks belong to
the mode $(2,2)$. The calculation is done in the Andreev
approximation.

\vspace{7mm}

\begin{figure}[ht]
\centerline{\epsfig{figure=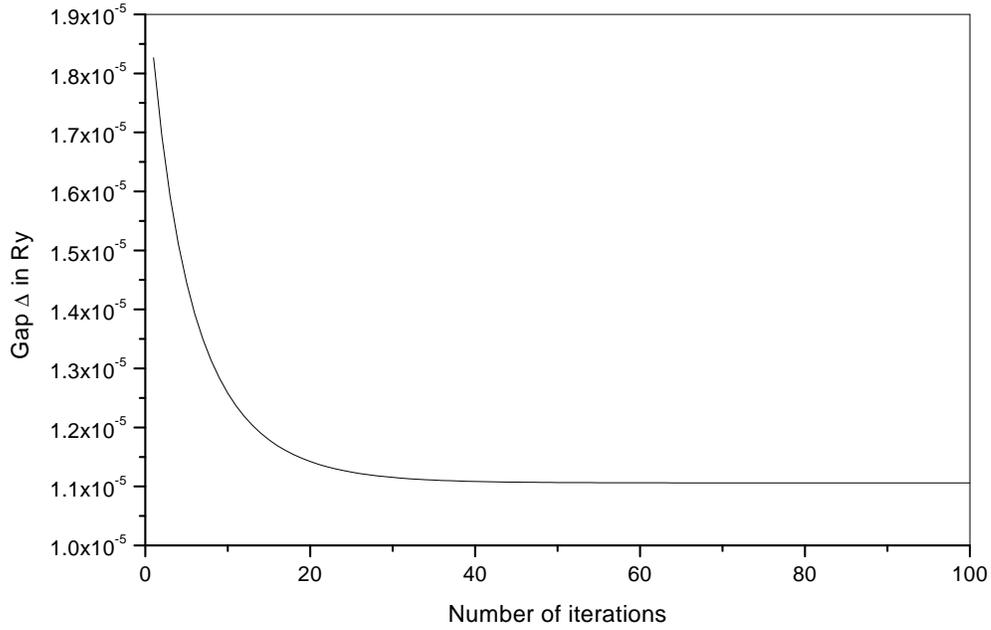,height=10.0cm}}
\caption[]{The gap against the number of iterations for a
bar-shaped superconductor. Note that the value of the gap
stabilizes as the number of iterations increases. For the system
considered, $L_t=1000$ Bohr and $T=0.6$ K.} \label{PGapite}
\end{figure}

FIG.~\ref{PGapite}. The gap against the number of iterations for a
bar-shaped superconductor. Note that the value of the gap
stabilizes as the number of iterations increases. For the system
considered, $L_t=1000$ Bohr and $T=0.6$ K.

\vspace{7mm}

\begin{figure}[ht]
\centerline{\epsfig{figure=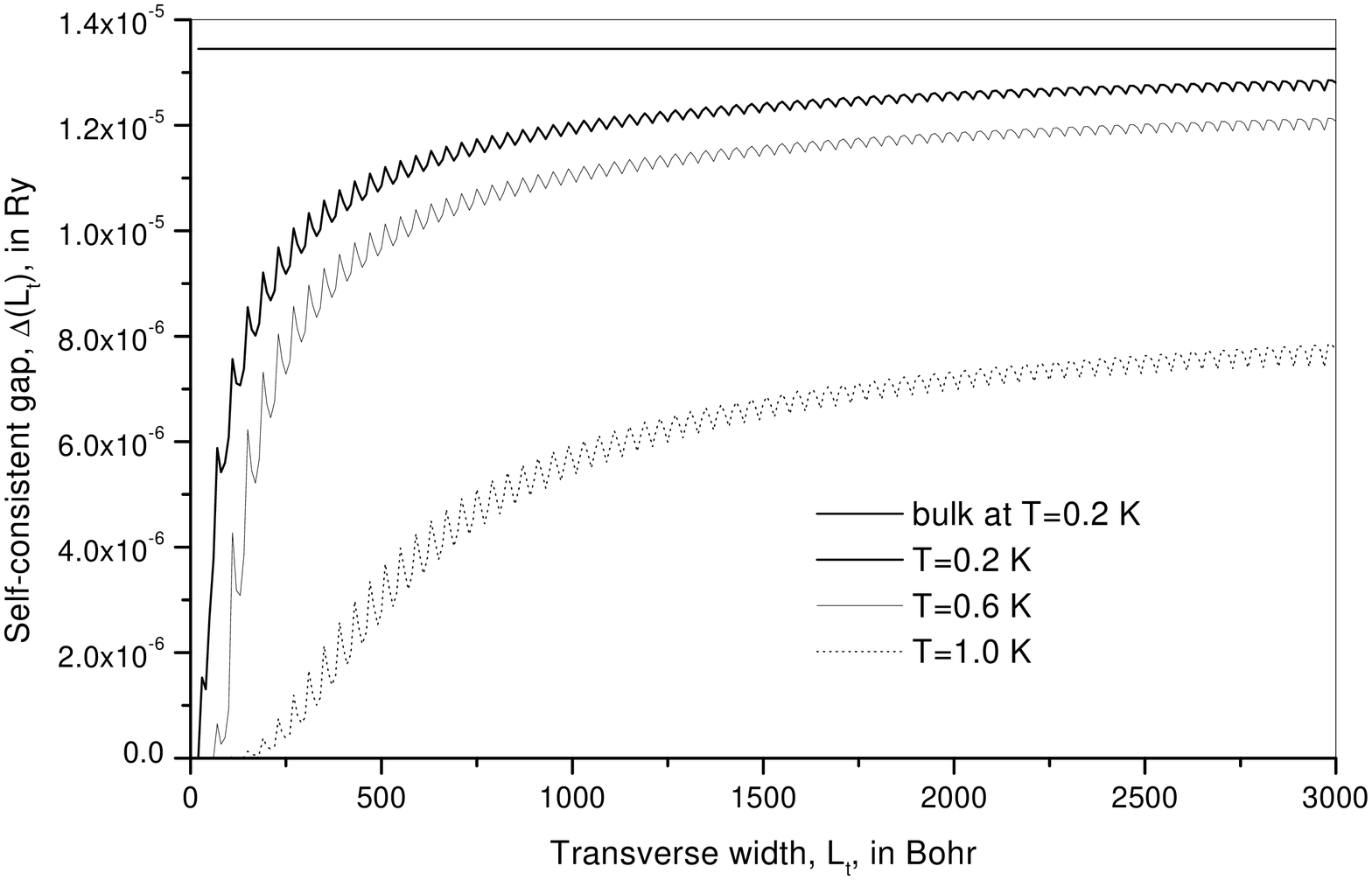,height=10.0cm}}
\caption[]{Plot of the self-consistent gap function against the
transverse length $L_t$ for a bar-shaped superconductor at
different temperatures. The number of iterations is 100.}
\label{PGapSLt}
\end{figure}

FIG.~\ref{PGapSLt}. Plot of the self-consistent gap function
against the transverse length $L_t$ for a bar-shaped
superconductor at different temperatures. The number of
iterations is 100.

\vspace{7mm}

\begin{figure}[ht]
\centerline{\epsfig{figure=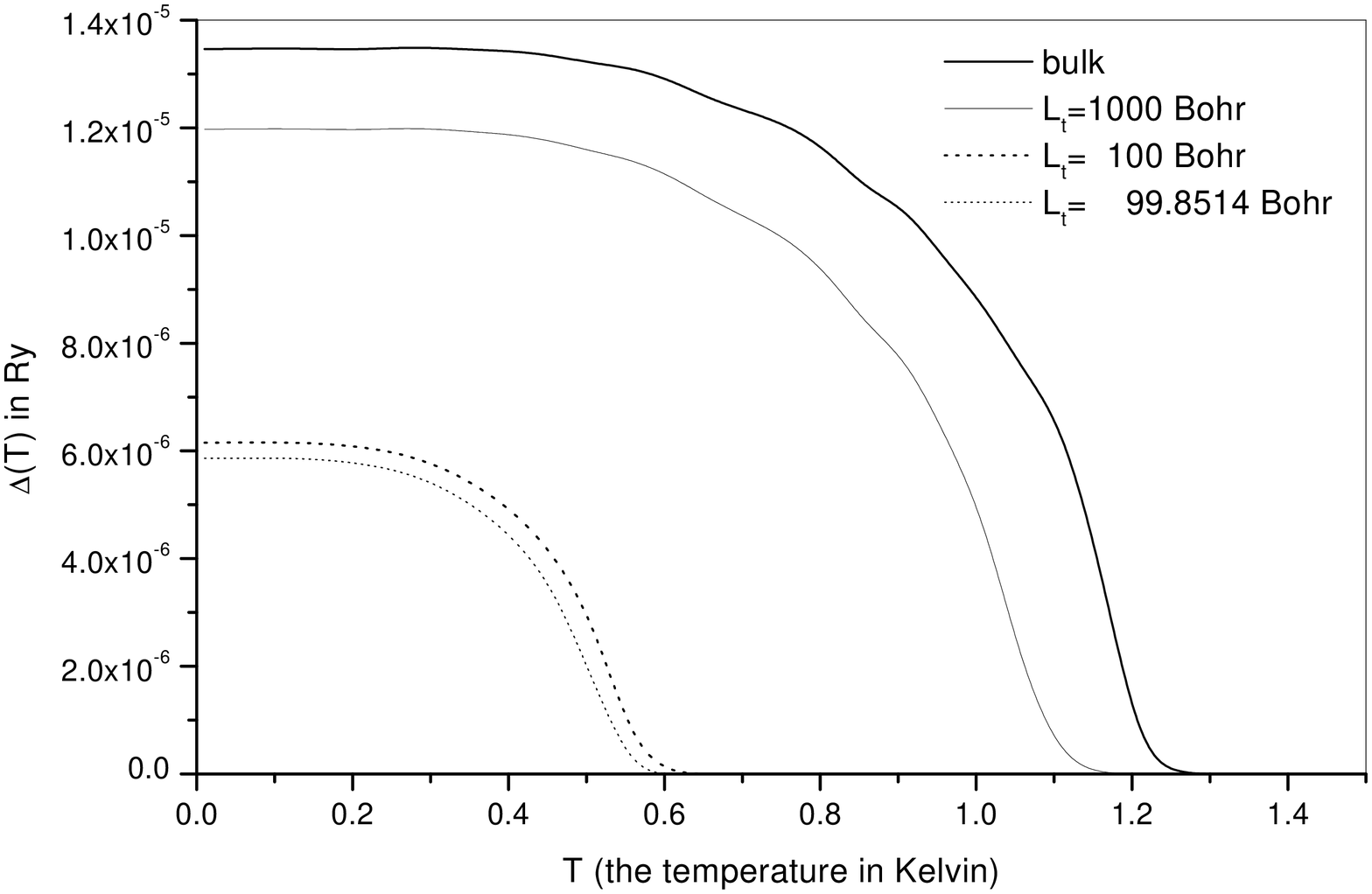,height=10.0cm}}
\caption[]{The temperature variation of the self-consistent gap
for different transverse widths.} \label{PGapST}
\end{figure}

FIG.~\ref{PGapST}. The temperature variation of the
self-consistent gap for different transverse widths.

\vspace{7mm}

\begin{figure}[ht]
\begin{center}
\setlength{\unitlength}{1cm}
\begin{picture}(18,18)
\put(3.5,1){\shortstack[c]{\epsfig{figure=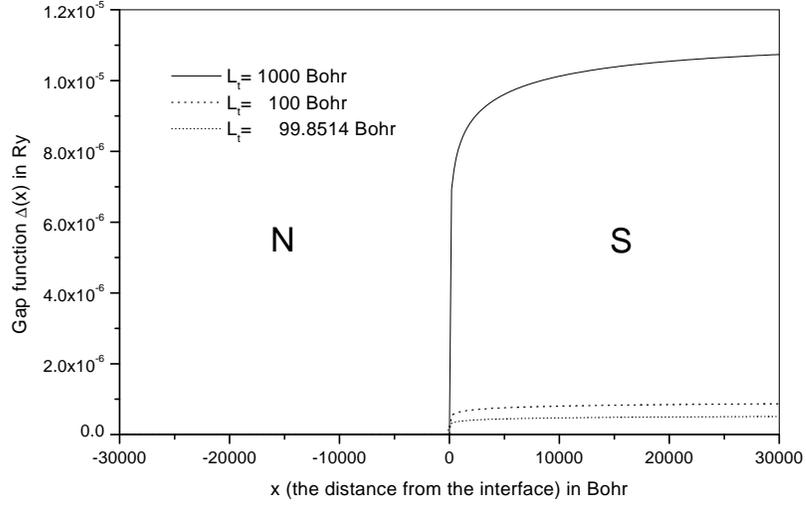,height=8.0cm}\\
\epsfig{figure=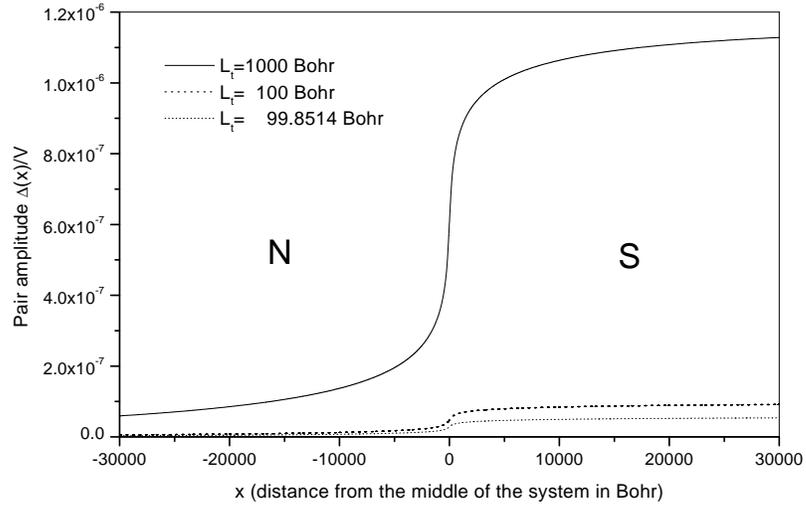,height=8.0cm}}} \put(9.3,8.8){(a)}
\put(9.3,0.5){(b)}
\end{picture}
\end{center}
\caption[]{(a) The gap
and (b) the pair amplitude against the distance from the
interface of a NS system at $T=0.6$ K. The interface is chosen at
$x=0$.}
\label{PGapNS}
\end{figure}

FIG.~\ref{PGapNS}. (a) The gap and (b) the pair amplitude against
the distance from the interface of a NS system at $T=0.6$ K. The
interface is chosen at $x=0$.

\vspace{7mm}

\begin{figure}[t]
\begin{center}
\setlength{\unitlength}{1cm}
\begin{picture}(18,18)
\put(3.5,1){\shortstack[c]{\epsfig{figure=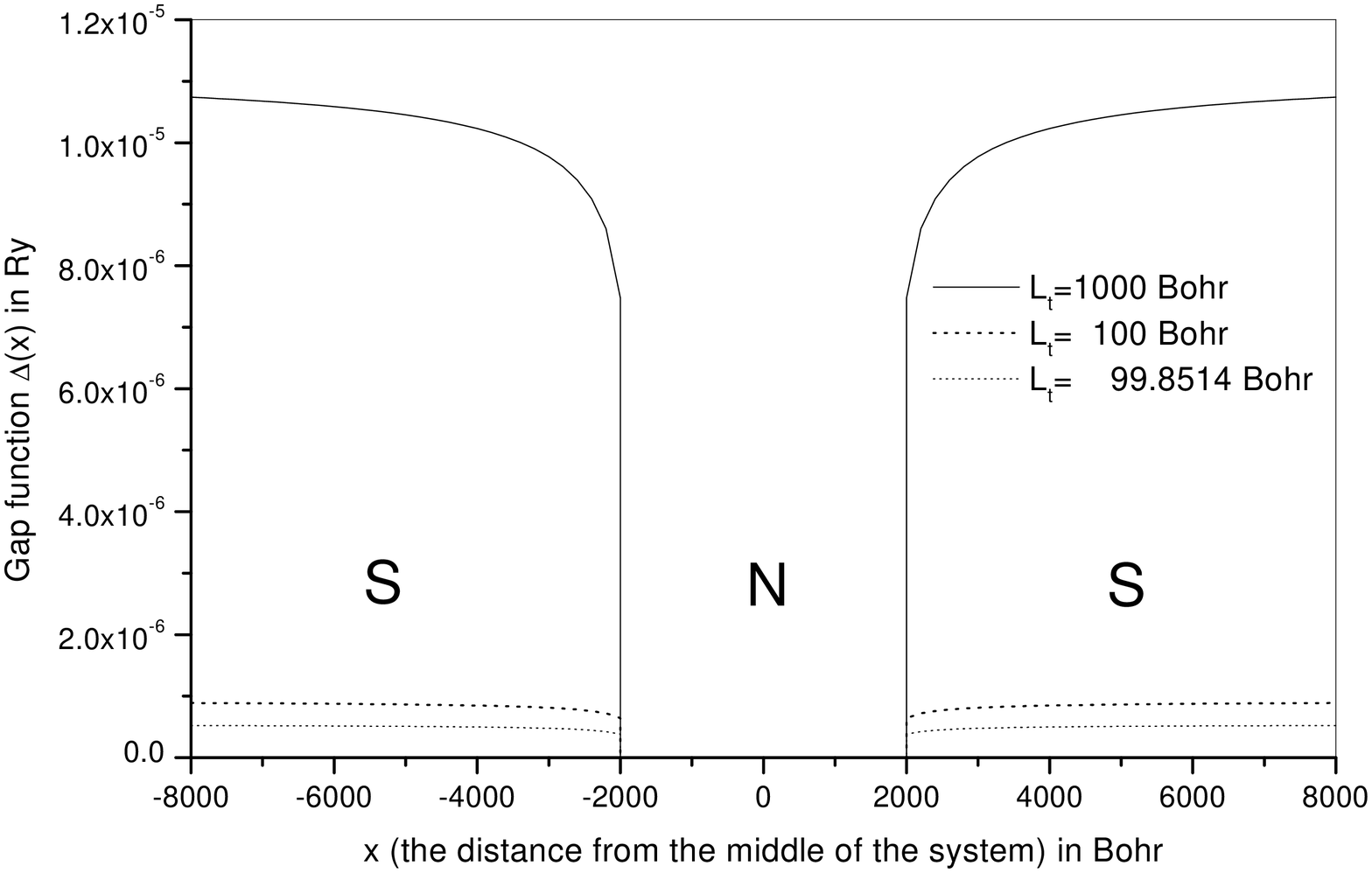,height=8.0cm}\\
\epsfig{figure=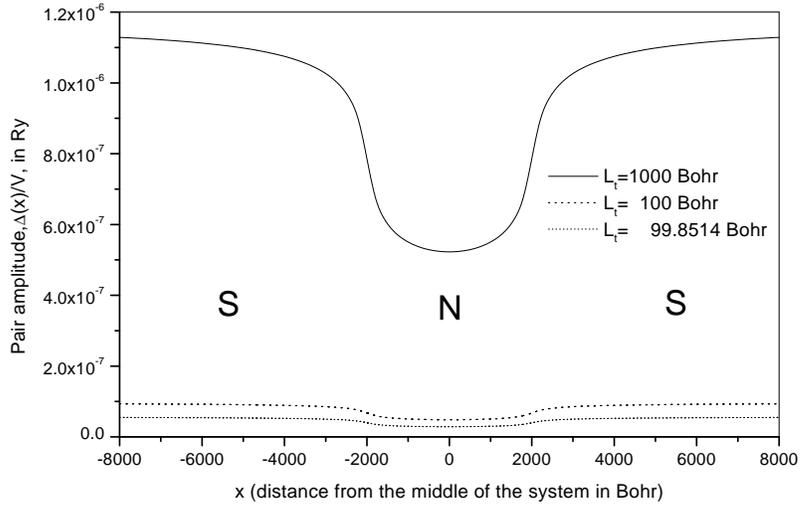,height=8.0cm}}} \put(9.3,8.8){(a)}
\put(9.3,0.5){(b)}
\end{picture}
\end{center}
\caption[]{(a) The gap function and (b) the pair amplitude of a
SNS system against the distance from the middle of the system
chosen at $x=0$. The interfaces are located at $x=\pm 2000$.}
\label{PGapSNS1}
\end{figure}

FIG.~\ref{PGapSNS1}. (a) The gap function and (b) the pair
amplitude of a SNS system against the distance from the middle of
the system chosen at $x=0$. The interfaces are located at $x=\pm
2000$.

\vspace{7mm}

\begin{figure}[t]
\centerline{\epsfig{figure=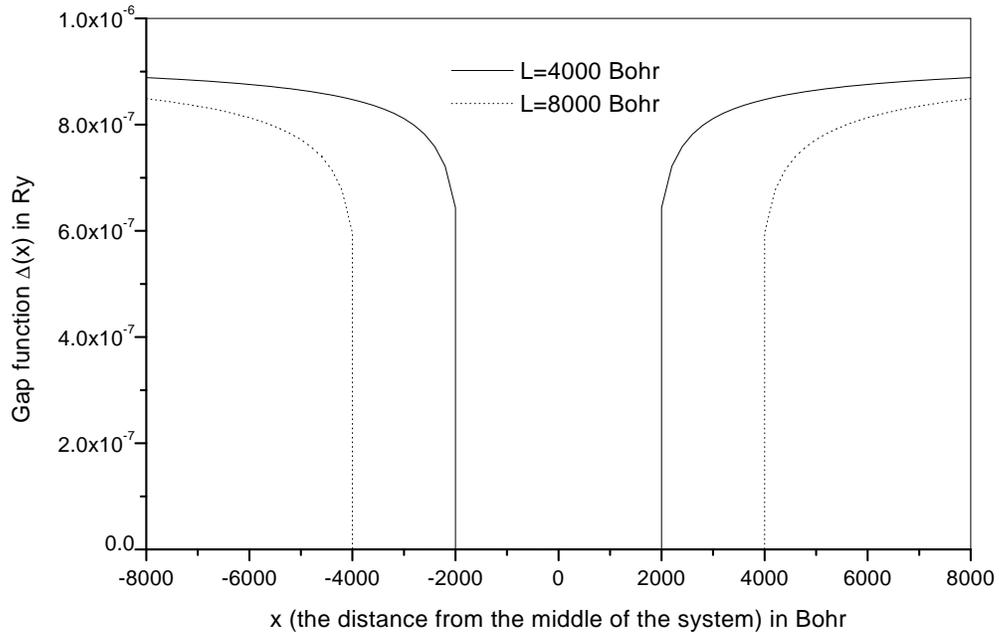,height=10.0cm}}
\caption[]{The gap function of a SNS system against the distance
from the middle of the system chosen at $x=0$ for different
values of $L$. The interfaces are located at $\pm 2000$ for
$L=4000$ Bohr and at $\pm 4000$ for $L=8000$ Bohr.}
\label{PGapSNS2}
\end{figure}

FIG.~\ref{PGapSNS2}. The gap function of a SNS system against the
distance from the middle of the system chosen at $x=0$ for
different values of $L$. The interfaces are located at $\pm 2000$
for $L=4000$ Bohr and at $\pm 4000$ for $L=8000$ Bohr.

\vspace{7mm}

\begin{figure}[ht]
\begin{center}
\setlength{\unitlength}{1cm}
\begin{picture}(18,18)
\put(3.5,1){\shortstack[c]{\epsfig{figure=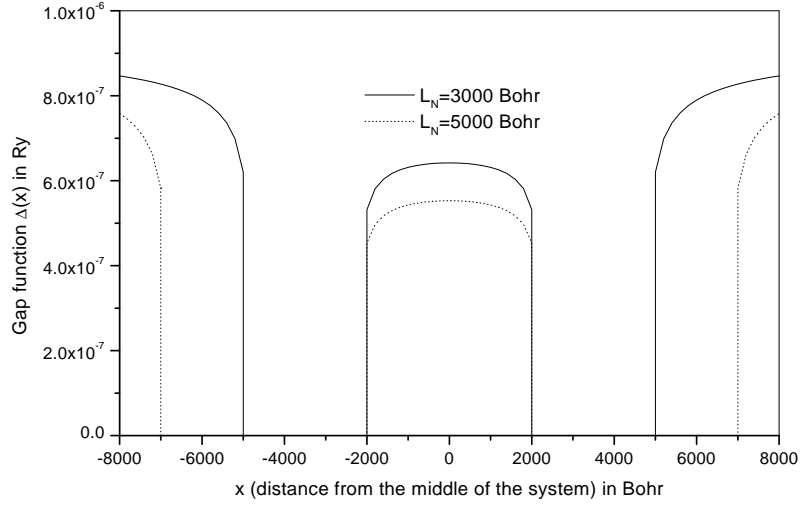,height=8.0cm}\\
\epsfig{figure=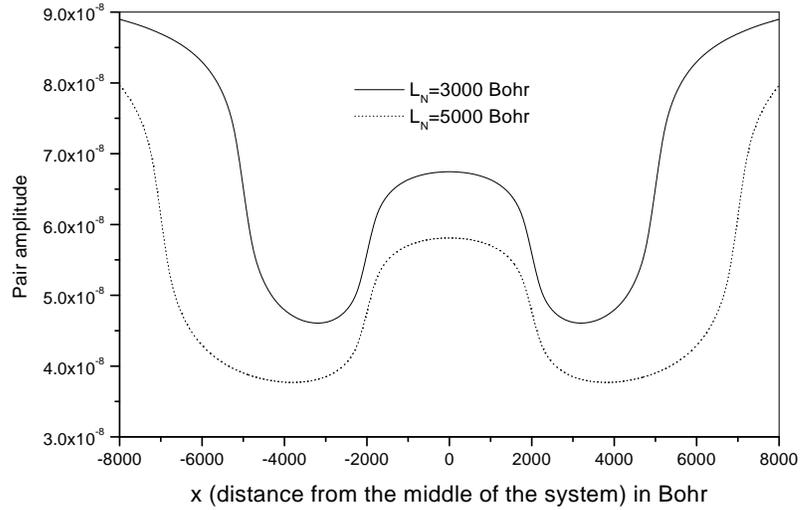,height=8.0cm}}} \put(9.3,8.8){(a)}
\put(9.3,0.5){(b)}
\end{picture}
\end{center}
\caption[]{(a) The gap function and (b) the pair amplitude against
the distance from the middle of a SNSNS system, which is chosen at
$x=0$. The interfaces are chosen at $x=\pm 2000, \pm 5000$ for
$L_N=3000$ Bohr (solid curve) and $x=\pm 2000, \pm 7000$ for
$L_N=5000$ Bohr (dotted curve). The transverse width is $L_t=100$
Bohr.} \label{PGapSNSNS1}
\end{figure}

FIG.~\ref{PGapSNSNS1}. (a) The gap function and (b) the pair
amplitude against the distance from the middle of a SNSNS system,
which is chosen at $x=0$. The interfaces are chosen at $x=\pm
2000, \pm 5000$ for $L_N=3000$ Bohr (solid curve) and $x=\pm
2000, \pm 7000$ for $L_N=5000$ Bohr (dotted curve). The
transverse width is $L_t=100$ Bohr.

\newpage

\end{document}